\documentclass[twocolumn]{custom_template}

\usepackage[english]{babel}
\usepackage[utf8]{inputenc}
\usepackage[T1]{fontenc}

\usepackage{graphicx}
\usepackage{combelow}
\usepackage{mathtools}
\usepackage{amsfonts, amssymb}
\usepackage{sidecap}

\usepackage{lipsum}

\usepackage[colorlinks,allcolors=black]{hyperref}

\title{\vspace{-1cm}\centering\bfseries\Large Theory of adhesion-driven self-organisation in growing tissues}

\author{\centering \normalsize Carles Falc\'o$^{1,*}$\;, Samuel W.~S. Johnson$^{1}$\;, Mohit P. Dalwadi$^{1}$\;, Philip K. Maini$^{1}$}

\date{}

\begin{document}

\twocolumn[\maketitle\par\vspace{-1.3cm}
\begin{center}
{$^{1}$Mathematical Institute, University of Oxford, OX2 6GG Oxford, United Kingdom}

{$^*$Correspondence: falcoigandia@maths.ox.ac.uk}
\end{center}
\centering\begin{minipage}{16cm} 
Cell invasion and spatial pattern formation are two distinct manifestations of cellular self-organisation in development, regeneration, and disease. Here, we develop and analyse a unified theoretical framework that links these two seemingly different behaviours within a single mechanistic model for adhesion-mediated self-organisation in growing cell populations. Using a multiscale analysis, we show that the balance between cell--cell adhesion, self-diffusion, and proliferation controls the emergence of distinct collective dynamics. We find that for weak adhesion, tissues invade through stable monotone fronts. As adhesion increases, invasion slows, fronts become unstable, leading to  aggregates and spatial patterns emerging behind the advancing edge. In two spatial dimensions, these instabilities generate fingering morphologies reminiscent of dysregulated invasion in cancer. Crucially, we show that density-dependent regulation of adhesion suppresses these instabilities and restores cohesive tissue expansion. Together, our results identify adhesion strength and its regulation as key determinants of whether tissues invade cohesively or fragment into patterns, and provide a unified framework for understanding collective migration, morphogenesis, and dysregulated growth.

\end{minipage}

\par\vspace{5ex}]


\section*{Introduction}
Cell invasion and spatial pattern formation are two fundamental manifestations of cellular self-organisation in development and disease, yet are generally studied as distinct processes within separate theoretical frameworks. During invasion, cells coordinate migration and proliferation to colonise available space, enabling for instance tissue morphogenesis, wound healing, and bacterial chemotaxis  \cite{scarpa2016collective, li2013collective, narla2021traveling}. Pattern formation similarly relies on the regulation of motility and proliferation; however, rather than colonising additional space, self-organisation generates spatial heterogeneity, giving rise to distinct regions populated by specific cell types. This capacity to spatially organise is central throughout embryonic development and in pathological contexts such as cancer \cite{wolpert1978pattern, muthuswamy2021self}. A natural question, then, is whether invasion and pattern formation can be understood as different dynamical regimes of a single underlying mechanism \cite{Ford2025PNAS}. 

\begin{figure*}[t!]
    \centering
\includegraphics[width=.98\linewidth]{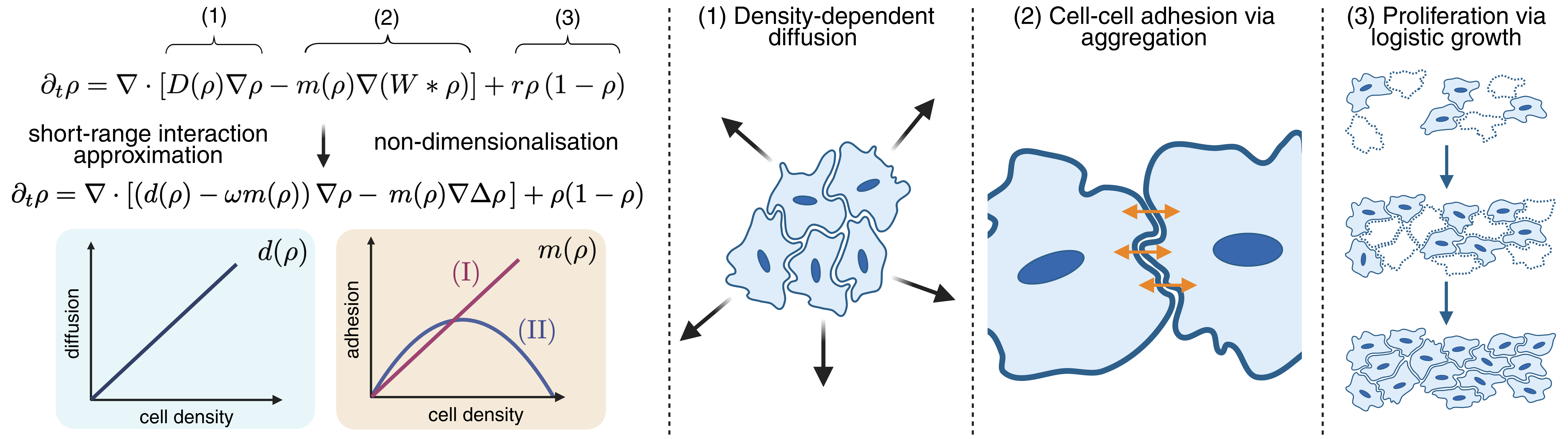}
    \caption{\textbf{Model schematic and interpretation.}
Left: continuum model showing the short-range interaction approximation from a non-local to a local formulation, with each numbered term corresponding to a distinct mechanism. Right: schematic representation of the three contributions: density-dependent self-diffusion, cell-cell adhesion, and density-modulated proliferation. Bottom left panels illustrate representative choices used in this work for the diffusion coefficient $d(\rho)$ and adhesion-based mobility $m(\rho)$; in particular, we show an unsaturated adhesion model (I) and a saturated adhesion model (II) that saturates at $\rho = 1$.}
    \label{fig: fig 1}
\end{figure*}

The mathematical concept of self-organisation in biology is often traced back to the seminal work of Turing \cite{turing1952chemical}. Turing's proposal that diffusion, traditionally associated with homogenisation, could instead drive the emergence of spatial patterns such as travelling waves, oscillations, stripes, and spots was both revolutionary and counter-intuitive \cite{green2015positional}. Since then, numerous developmental processes have been interpreted through the framework of so-called Turing instabilities, which arise in reaction–diffusion systems through the interplay of local activation and long-range inhibition of at least two morphogens. Such mechanisms have been invoked to explain periodic and spatially ordered structures across a range of developmental contexts, including for instance epithelial ridges, skin appendages, and digit patterning~\cite{economou2012periodic, lin2009spots, sick2006wnt, raspopovic2014digit}.

Despite its explanatory power in many biological systems, the classical Turing framework is not always directly applicable, as it requires at least two interacting species---a local activator and a long-range inhibitor---to generate spatial structure. In many biological contexts, only one relevant species can be identified experimentally, or the diffusivity ratios in multispecies systems lie outside the ranges required for robust pattern formation. In addition, classical Turing mechanisms typically select patterns with a characteristic intrinsic length scale, raising questions about robustness \cite{maini2012turing} and scaling \cite{ben2011scaling} during development, where tissues and organisms grow substantially while spatial organisation must be maintained. These limitations have motivated alternative theoretical frameworks that likewise invoke \emph{competing mechanisms} acting on distinct length scales. One such example is mechanochemical patterning, in which tissue mechanics and geometry can effectively provide long-range inhibitory feedback on chemical gradients. Through this coupling, self-organisation can emerge even from relatively simple interaction rules \cite{brinkmann2018post} and give rise to more generic instabilities than those predicted by classical two-morphogen Turing systems \cite{recho2019theory}.

Another paradigm based on the idea of competing mechanisms at different length scales is that of adhesion-based aggregation. Differential adhesion has long been recognised as a mechanism for tissue self-organisation in mixtures of two cell populations \cite{foty2005differential,tsai2022adhesion}. However, even in single-species systems, adhesion can generate spatial structure, typically in the form of spot-like aggregates \cite{ArmstrongPainterSherratt, campbell2021cooperation}. In this case, volume exclusion or contact inhibition of locomotion result in localised dispersal, while adhesion acts over longer length scales to promote aggregation. Mathematically, these effects are typically represented by non-local equations that encode long-range adhesive interactions, following the widely used model of Armstrong et al.~\cite{ArmstrongPainterSherratt} and subsequent extensions~\cite{MurakawaTogashi, CarrilloMurakawaCellAdhesion}. More recently, the single-species model has been recast within the broader framework of \emph{directed cell migration}, encompassing non-local mechanisms such as chemotaxis, and shown to provide a robust route to pattern formation and morphogenesis \cite{yu2025directed}. In parallel, asymptotic approaches exploiting the small interaction length scale relative to tissue size have yielded local approximations of these non-local models, while preserving their capacity to generate single-species aggregation and multi-species sorting via differential adhesion \cite{falco2022local,falco2025nonlocal}. 

Here we present a comprehensive theory of adhesion-based self-organisation in growing and migrating cell populations. Building on recent results obtained via a short-range interaction approximation, we derive an analytically tractable model for migrating cells in which density-dependent self-diffusion acts locally, while adhesion promotes aggregation over longer length scales. The resulting framework bears a formal resemblance to Cahn--Hilliard-type equations with proliferation \cite{khain2008generalized, khain2007role}, although it arises here through a systematic coarse-graining of cell-level interactions and is governed by a single effective parameter capturing the balance between adhesion, self-diffusion, and proliferation. We find that when adhesion is weak, the model predicts robust cell invasion in the form of travelling waves, which we characterise analytically using a multiscale analysis. For stronger adhesion, invasion occurs behind a wave of spot formation. The shape and characteristic length scale of these patterns can likewise be determined analytically. In the transition regime between these behaviours, we uncover non-monotonic invasion fronts in one spatial dimension. Extending the analysis to two spatial dimensions reveals that these fronts are unstable to transverse perturbations, leading to fingering instabilities of initially planar waves. This behaviour closely resembles the dysregulated invasion patterns often observed in cancer progression and tumour growth.
Finally, we demonstrate that when adhesion is regulated in a density-dependent manner, these instabilities can be suppressed, restoring robust invasion even in the strong-adhesion regime.

Taken together, our results show that invasion and pattern formation need not be regarded as distinct processes, but can instead emerge as different dynamical regimes of the same underlying mechanism of cell--cell adhesion.

\section*{Results}
\subsection*{Modelling adhesive interactions}

We begin with a widely used model of cell migration that incorporates adhesive interactions through a non-local advection term (see Fig.~\ref{fig: fig 1} for a schematic). Variants of this formulation are common in the literature \cite{ArmstrongPainterSherratt,MurakawaTogashi,CarrilloMurakawaCellAdhesion,chenpainter2020nonlocal}, and a unifying single-species framework has recently been proposed by Yu et al. \cite{yu2025directed}.

Cells are described by a spatio-temporal density, $\rho(\mathbf{x}, t)$, where $\mathbf{x}\in\Omega\subset\mathbb{R}^n$, $n \in \{1,2\}$, and $t\geq0$. Unless stated otherwise, we consider domains, $\Omega = [-L/2,L/2]^n$, with $L$ large relative to the range of adhesive interactions and periodic boundary conditions on cell density. 
Cell movement arises from two mechanisms. Firstly, self-diffusion, representing unbiased stochastic motion and/or volume exclusion effects, is described by a density-dependent diffusivity $D(\rho) \ge 0$. Secondly, adhesion drives attraction between neighbouring cells via a spatial convolution, $(W * \rho)(\mathbf{x},t)  = \int_\Omega W\left(\mathbf{x} - \mathbf{y}\right)\rho(\mathbf{y},t)\,\mathrm{d}\mathbf{y}$, where $W(\cdot)$ is an attractive interaction potential encoding the strength and range of adhesion. We further assume that adhesion depends on cell density through an adhesion-based mobility coefficient $m(\rho) \ge 0$. Finally, cell proliferation is modelled by logistic growth with an intrinsic rate $r > 0$ and a saturation density at $\rho = 1$. Combining these effects yields the following non-local reaction--aggregation--diffusion equation:
\begin{equation}
    \partial_t \rho
    = \nabla \cdot \left[ D(\rho) \nabla \rho-m(\rho) \nabla (W * \rho) \right]
    + r \rho \left(1 - {\rho}\right).
    \label{eq:nonlocal_model}
\end{equation}
Modelling assumptions regarding how adhesion depends on local cell density are encoded in the choice of the mobility function, $m(\rho)$. In particular, it is natural to require $m(0)=0$, since isolated cells cannot experience adhesive interactions. A simple and widely used choice is $m(\rho)=\rho$, for which the adhesive flux scales linearly with cell density. In this case, each cell contributes equally to adhesion independently of local crowding; we refer to this as the \emph{unsaturated adhesion model} (model~$(\mathrm{I})$). Alternatively, density-dependent forms such as $m(\rho)=\rho(1-\rho)$ reflect the reduction of effective motility in highly crowded environments, where frequent cell--cell contacts hinder movement. This choice defines the \emph{saturated adhesion model} (model~$(\mathrm{II})$).

While the non-local model in Eq.~\eqref{eq:nonlocal_model} has been widely successful in explaining classical cell-sorting patterns arising from differential adhesion, the non-local nature of the model hinders subsequent analysis. To address this, we consider a short-range interaction limit~\cite{BernoffTopazCH,falco2022local, falco2025nonlocal} and derive a local model of cell migration that still accounts for adhesive interactions. Such a model has been shown to preserve differential adhesion patterns in the absence of proliferation and under density-independent adhesion (i.e.\ when $m(\rho) = \rho$). We note that, in the absence of proliferation, the resulting local formulation shares structural features with thin-film-type equations, though here, the sign of the second-order term promotes aggregation rather than spreading; see~\cite{evans2007source} for related asymptotic analyses of thin-film models. In Appendix~\ref{sec:appendix-nonlocal-to-local}, we extend the nonlocal-to-local derivation to the general case of density-dependent adhesion and include proliferation, obtaining the following local model:
\begin{equation*}
    \partial_t\rho = \nabla\cdot\left[\left(D(\rho) -\omega_0 m(\rho)\right)\nabla\rho - \frac{\omega_2}{2n}\,{m}(\rho)\nabla\Delta\rho\,\right] + r\rho(1 - \rho)\,,
\end{equation*}
 where $\omega_j$ denotes the $j$-th order moment of the potential $W$. For simplicity, we rescale time and space using the change of variables $\Tilde{t} = r t$, and $\Tilde{\mathbf{x}}=(2nr/\omega_2)^{1/4}\mathbf{x}$. Then, omitting the tildes, we obtain the local model:
\begin{equation}
     \partial_t\rho = \nabla\cdot\left[\left(d(\rho) -\omega m(\rho)\right)\nabla\rho - \,{m}(\rho)\nabla\Delta\rho\,\right] + \rho(1 - \rho)\,,
     \label{eq: local model general}
\end{equation}
where
\begin{equation}
    d(\rho) =  \sqrt{\frac{2n}{r\omega_2}}D(\rho)\,,\qquad\omega = \sqrt{\frac{2n}{r\omega_2}}\omega_0\,.\label{eq: nondimensional}
\end{equation}

Motile cells frequently reorient their velocities upon contact, a behaviour known as contact inhibition of locomotion~\cite{alert2020physical}. This effect can be modelled by self-diffusion arising from localised cell–cell repulsive interactions. Systems with short-range repulsive interactions have been extensively studied in the context of hard~\cite{bruna2012excluded} and soft-sphere models~\cite{bruna2017diffusion} and on-lattice processes~\cite{bakerAspectRatio2, DysonVolumeExclusion, falco2022random}, where they typically give rise to nonlinear diffusion of the porous-medium type. In this spirit, and unless otherwise stated, we adopt a density-dependent diffusivity of the form $d(\rho) = \alpha\rho$, where $\alpha>0$ quantifies the strength of repulsive interactions between cells, or equivalently, the strength of contact inhibition of locomotion. We note that the saturation density $\rho=1$ represents a proliferative carrying capacity rather than a physical jamming threshold; self-diffusion remains well-defined at and above this density, consistent with the assumption that cells can transiently exceed the carrying capacity before growth regulation brings the density back toward saturation.

%this section admits some reordering, and maybe redefining m(\rho)

% Some ideas for subsections
\subsection*{Adhesion-based pattern formation}

\begin{figure}[b!]
    \centering
\includegraphics[width=0.9\linewidth]{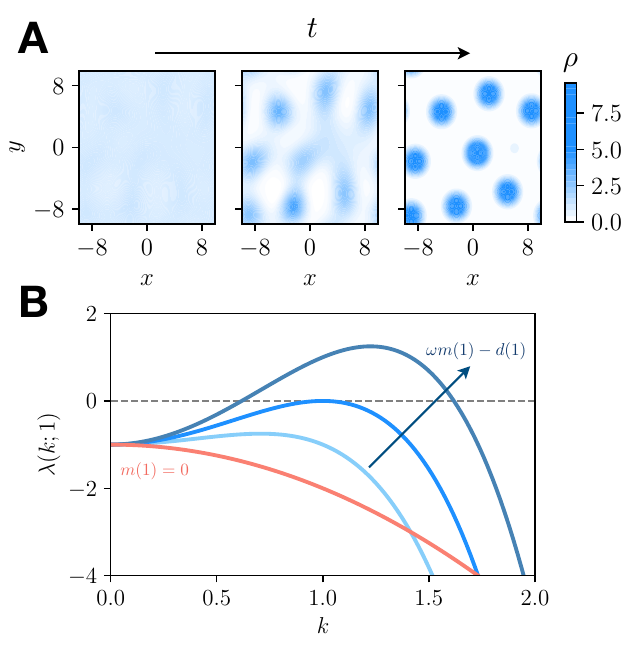}
   \caption{
\textbf{Adhesion-based patterning. }
\textbf{(A)} Example of pattern formation in the conservative model without proliferation, illustrating the emergence of spatial structure driven purely by adhesion. Functions are chosen as $m(\rho)=\rho$, $d(\rho)=\alpha\rho$, with $\alpha-\omega=-2$, so that the pattern formation condition $H''(\phi)<0$ is satisfied. 
\textbf{(B)} Dispersion relation $\lambda(k;1)$ at the saturated state $\phi=1$ for the model with logistic growth. Blue curves correspond to increasing adhesion strength. The red curve corresponds to $m(1)=0$, representing saturating adhesion at high densities; in this case pattern formation is suppressed.
}
    \label{fig: fig 2}
\end{figure}
In the absence of proliferation, the model has a variational structure with respect to a weighted Wasserstein metric \cite{antonio2024competing} and is mass-conservative. The inclusion of logistic growth breaks this structure and renders the system non-conservative, a modification which generates travelling waves and spatio-temporal chaos. In particular, the model can be rewritten in the form (see Appendix~\ref{sec:appendix-energy} and~\cite{falco2022local,falco2025nonlocal}):
\begin{equation*}
    \partial_t\rho = \nabla\cdot\left(m(\rho)\,\nabla\frac{\delta\mathcal{F}}{\delta \rho}\right) + \rho(1-\rho)
\end{equation*}
with respect to the free energy 
\begin{equation*}
    \mathcal{F}[\rho] = \int_\Omega \left(\frac{1}{2}|\nabla\rho|^2+H(\rho)\right)\,\mathrm{d}\mathbf{x}\,,
\end{equation*}
where $H(\rho)$ satisfies $H''(\rho) = d(\rho)/m(\rho) - \omega$. The associated energy formulation provides a natural framework within which to understand the emergence of patterns and the structure of steady states, when they exist, as these correspond to stationary points of the free energy functional, satisfying: $\delta\mathcal{F}/\delta\rho = -\Delta\rho + H'(\rho)=C$, for a constant $C$. In this formulation, the functional $\int_\Omega H(\rho)\,\mathrm{d}\mathbf{x}$ plays the role of a \emph{negative entropy}, in that it promotes aggregation and, when sufficiently dominant, can drive pattern formation. In the conservative model, the onset of pattern formation is governed by the convexity of $H$. Specifically, any homogeneous state $\rho(\mathbf{x},t)=\phi$ becomes linearly unstable whenever $H''(\phi)<0$ (i.e.\ when adhesion is strong enough): $\omega>{d(\phi)}/{m(\phi)}$---see Fig~\ref{fig: fig 2}A for a pattern formation example where $d(\rho), \,m(\rho)\propto \rho$.

With logistic growth, the reaction term $\rho(1-\rho)$ admits exactly two homogeneous steady states, $\phi=0$ and $\phi=1$. Their stability is determined by the dispersion relation $\lambda(k;\phi)$ obtained by linearising about $\rho=\phi$ (see Appendix~\ref{sec:linear-stability} for details). Near $\phi=0$, growth is destabilising, and thus this state is always linearly unstable.

Close to $\phi=1$, growth is stabilising, that is, it relaxes perturbations back to the saturation density. Instability can only arise if adhesion sufficiently dominates diffusion to create growing spatial modes despite the baseline decay; this yields the dispersion relation\begin{equation*}
    \lambda(k;1) = -1 -(d(1)-\omega m(1))k^2-m(1)k^4\,,\label{eq: main lambda la}
\end{equation*}and condition~\eqref{eq: B}:
\begin{equation}
     (\rho = 1 \quad \text{unstable}) \quad   \omega m(1) -d(1)  > 2\sqrt{m(1)}\,.\tag{$\dagger$} \label{eq: B}
\end{equation}
Condition~\eqref{eq: B} (see Fig.~\ref{fig: fig 2}B) highlights that pattern formation  is controlled by three factors. Firstly, $\omega$ sets the effective strength of adhesion relative to proliferation---larger $\omega$ makes it easier for spatial modes to overcome the local restoring tendency. Secondly, the density-dependence of adhesion near saturation enters through $m(1)$; if $m(1)$ is small, the patterning threshold is increased. Thirdly, diffusion near saturation competes directly with adhesion via $d(1)$; larger $d(1)$ stabilises the homogeneous state.

\subsection*{Cell invasion and the role of adhesion}

\begin{figure*}[t!]
    \centering
\includegraphics[width=.9\linewidth]{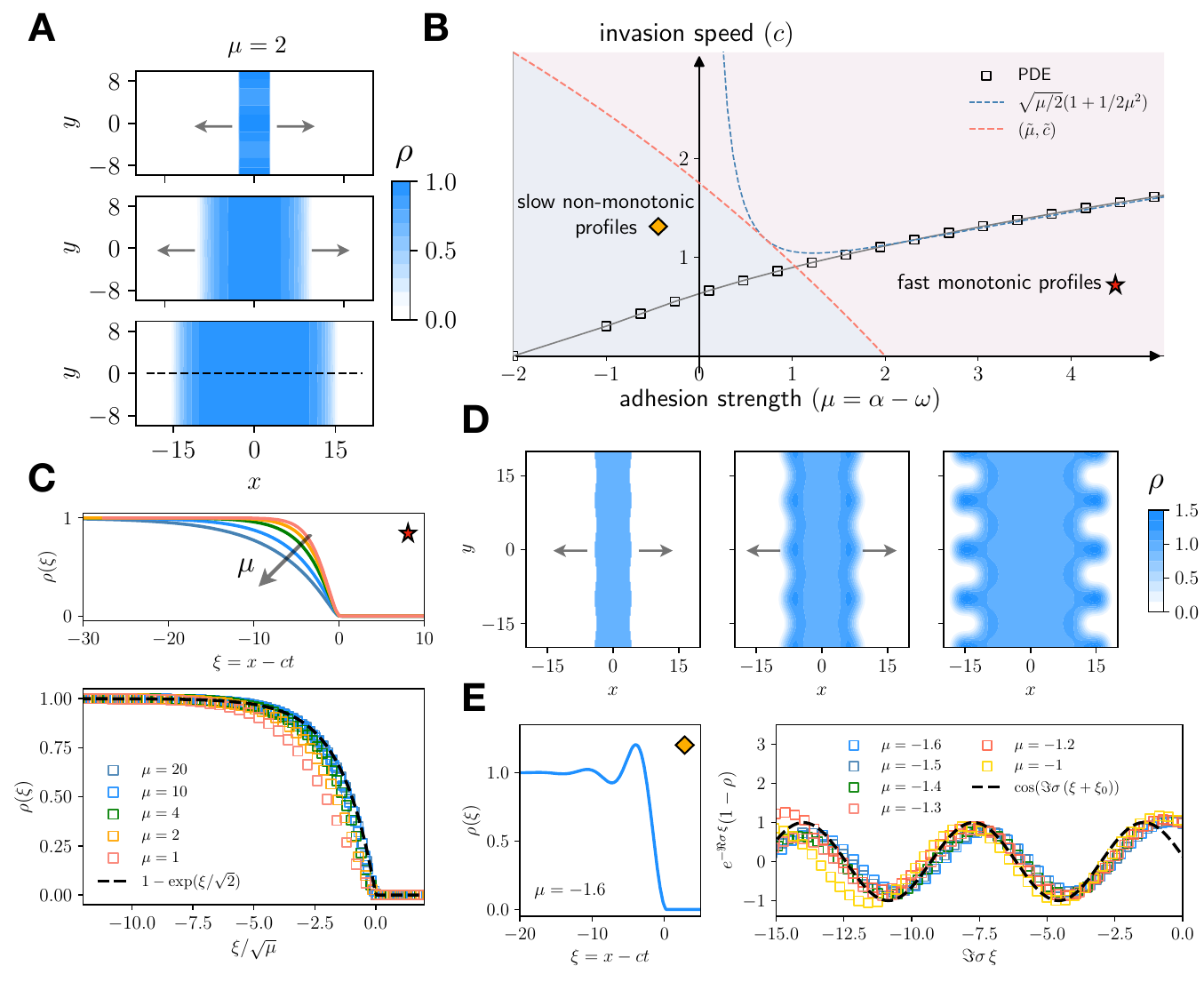}
    \caption{\textbf{Cell invasion in the unsaturated adhesion model} (Eq.~\eqref{eq: local model constant}). (A) Two-dimensional numerical simulations in the weak-adhesion regime ($\mu=2$) show a robust planar invasion front propagating symmetrically into the empty region, with a monotonic profile across the interface (black dashed line). (B) Selected invasion speed $c$ versus adhesion strength $\mu$: Eq.~\eqref{eq: local model constant} measurements (squares) are compared with the large-$\mu$ asymptotic prediction (blue dashed line, Eq.~\eqref{eq: speed large mu}); the red dashed curve marks the onset of oscillatory tails in the $(\mu,c)$ plane, separating fast monotonic profiles from slow non-monotonic profiles, and its intersection with the selected speed gives the critical value $\mu_c$. (C) One-dimensional travelling-wave profiles $\rho(\xi)$ in the weak adhesion regime, together with collapse of weak-adhesion profiles under the rescaling $\xi/\sqrt{\mu}$ and agreement with the outer asymptotic approximation (bottom, Eq.~\eqref{eq: constant weak adhesion}). (D) Near the patterning threshold ($\mu\to-2^+$), fronts develop damped oscillatory tails (left); after rescaling by the decay length scale, the far-field oscillations agree with the cosine prediction from the linearised analysis (right, Eq.~\eqref{eq: sigma} with $\xi_0\approx 14$ and $\Im\sigma\approx \sqrt{2-\mu}/2$). (E) In two spatial dimensions, planar fronts can become transversely unstable for stronger adhesion, so small perturbations grow into localised protrusions that lead the invasion front.}
    \label{fig: fig 3}
\end{figure*}

\begin{SCfigure*}[.31][t!]
\begin{wide}
    \includegraphics[width = .7\textwidth]{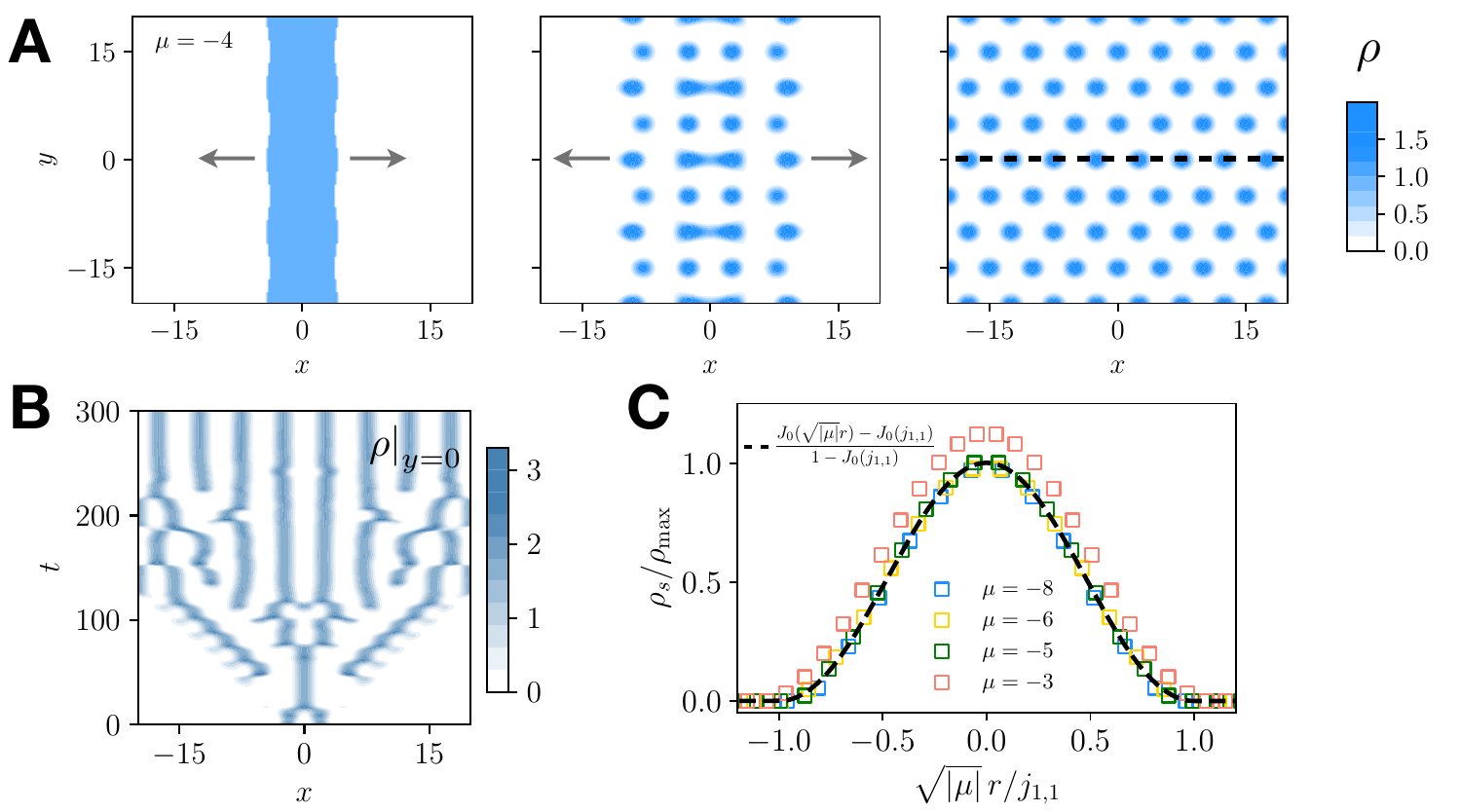}
    \caption{\textbf{Strong adhesion: invasion and patterning $(\mu < -2)$. } (A) A planar invading front destabilises and gives rise to a periodic array of localised aggregates behind the invasion front ($\mu=-4$).
(B) Time evolution of $\rho|{y=0}$, showing invasion and subsequent pattern formation (blacked dashed line in (A)).
(C) Rescaled numerical aggregate profiles for different values of the adhesion strength, $\mu$, compared with the asymptotic prediction in Eq.~\eqref{eq: steady state mu}; the agreement remains strong even near the threshold $\mu=-2$.}
    \label{fig: fig 4}
    \end{wide}
\end{SCfigure*}

Next we focus on a model arising from a simple choice of adhesion-based mobility. We assume each cell contributes equally to the population adhesive strength, yielding a linear dependence $m(\rho) = \rho$. We refer to this as the \emph{unsaturated adhesion} model or model~\eqref{eq: local model constant}. For a linear diffusion coefficient, $d(\rho) = \alpha\rho$, Eq.~\eqref{eq: local model general} reads
\begin{equation}
    \partial_t\rho = -\nabla\cdot\left[\rho\,\nabla\left(\Delta\rho - \mu\rho\right)\right] + \rho(1-\rho)\,,
     \label{eq: local model constant}\tag{$\mathrm{I}$}
\end{equation} \setcounter{equation}{0}    
 where the only parameter, $\mu = d(1) - \omega m(1)= \alpha - \omega$,  represents the ratio between net adhesion (adhesion minus self-diffusion) and proliferation. 
Positive values of $\mu$ indicate that diffusive interactions dominate over adhesion and proliferation. 
Values of $\mu$ near zero correspond to a balance between repulsion and adhesion, or to cases where proliferation is comparatively strong. 
Negative values of $\mu$ signify that adhesion dominates, promoting cell clustering. The parameter $\mu$ also defines the onset of pattern formation according to condition~\eqref{eq: B}, at $\mu = -2$. We focus first on the regime where the steady state $\phi =1$ is stable ($\mu>-2$) and cell invasion is expected.

In the regime of weak adhesion, corresponding to large positive $\mu$, numerical simulations reveal a robust invasion process characterised by a planar, monotonic travelling front connecting the unstable state $\rho=0$ to the saturated state $\rho=1$ (Fig.~\ref{fig: fig 3}A). Moreover, the invasion speed converges to that of a classical reaction–diffusion system with a linear-in-density diffusivity, i.e. the Porous-Fisher model~\cite{murray2001mathematical, falco2023quantifying}.
This behaviour can be made precise by looking at one spatial dimension and analysing travelling wave solutions of~\eqref{eq: local model constant} in the limit $\mu\to+\infty$. Writing $\rho(x,t)=\rho(\xi)$ with $\xi=x-ct$, and pinning the point at which $\rho$ first vanishes to be $\xi=0$ (exploiting the translational invariance of the travelling-wave equation), an analysis using the method of matched asymptotic expansions yields the composite approximation
\begin{align}
    \rho(x,t) & \sim \begin{cases}
         1 -e^{\xi/\sqrt{2\mu}},&\!\mbox{ for }\,|\xi/\sqrt{\mu}| = \mathcal{O}(1)\,;
         \\
          (e^{\sqrt{\mu}\xi}-\sqrt{\mu}\xi - 1)/\sqrt 2 \mu,&\!\mbox{ for } \,|\xi\sqrt{\mu}| = \mathcal{O}(1)\,;
    \end{cases}
    \label{eq: constant weak adhesion}
\end{align}
for $\xi \leq 0$, and $\rho = 0$ for $\xi >0$.  The asymptotic wave speed reads
\begin{equation}
c \sim \sqrt{\frac{\mu}{2}} + \frac{1}{2\sqrt{2}\mu^{3/2}}.\label{eq: speed large mu}
\end{equation}
Unlike in previously studied Cahn–Hilliard-type equations with proliferation, these invasion fronts are \emph{pushed} rather than \emph{pulled}, in the sense that their speed is selected by the nonlinear bulk dynamics rather than by the linear dynamics at the leading edge \cite{khain2008generalized}. Details of the derivation are given in Appendix~\ref{sec: constant appendix}. We observe excellent agreement between the asymptotic theory and numerical simulations (Fig.~\ref{fig: fig 3}B--C). For completeness, in Appendix~\ref{sec: finite contact angle} we also derive fully explicit travelling wave solutions with a finite contact angle as the density approaches zero. For completeness, in Appendix~\ref{sec: finite contact angle} we also derive fully explicit travelling-wave solutions with a non-zero contact angle as the density approaches zero. While these solutions solve the model in the comoving frame, they have not been observed in time-dependent simulations.

\subsection*{Intermediate  adhesion induces transverse instabilities and non-monotonic fronts}

As adhesion strength increases (that is, as $\mu$ decreases), planar fronts in two spatial dimensions can become unstable to transverse perturbations, unlike the robust invasion pattern observed for weak adhesion (Fig.~\ref{fig: fig 3}A). In particular, small deviations from a perfectly planar front can cause the cell density to aggregate into localised tips near the front, which then drive the invasion through protrusions or fingers (Fig.~\ref{fig: fig 3}D). While we leave a formal analysis of this instability for future work, we expect it to arise from the competition between adhesion and diffusion at the front, with proliferation continuing to sustain the invasion. We also do not expect every perturbation to cause instability, so planar fronts may still form.

To better understand how intermediate adhesion changes front behaviour, we return to the one-dimensional problem and analyse the structure of travelling waves. Although the transverse instability is inherently two-dimensional, the one-dimensional analysis reveals a related change in front structure: invasion waves can lose monotonicity once adhesion is sufficiently strong, that is, when $-2<\mu \leq \mu_c$. Such non-monotonic fronts are reminiscent of expansion profiles observed in epithelial (MDCK) monolayers, which are known to exhibit collective motion mediated by strong cell–cell adhesions \cite{heinrich2020size, falco2023quantifying}.
To characterise the onset of this behaviour, we examine the dynamics at the back of the travelling front, where the density is close to saturation. Linearising about this state allows us to quantify the growth of small perturbations. For weak adhesion, all perturbations decay monotonically; however, below a critical value of $\mu\leq \mu_c$, the leading eigenvalues become complex, indicating oscillatory growth and the emergence of non-monotonic front profiles. This transition defines an explicit curve in the $(\mu,c)$ plane, shown in Fig.~\ref{fig: fig 3}B; see Appendix~\ref{sec: travelling fronts} for the explicit expression. The threshold for non-monotonicity is obtained by intersecting this curve with the selected wave speed. Using the leading-order asymptotic approximation for $c(\mu)$ (Eq.~\eqref{eq: speed large mu}), we estimate the critical value to be $\mu_c \approx 1.18$, which shows good qualitative agreement with Fig.~\ref{fig: fig 3}B.

Closer to the pattern formation threshold, $\mu = -2$, we can further characterise the dynamics of these non-monotonic one-dimensional travelling fronts. A linearisation of Eq.~\eqref{eq: local model constant}, $\rho(\xi)-1\propto e^{\sigma \xi}$, allows us to quantify how density profiles deviate from saturation based on $\sigma$. In this regime, the selected wave speed vanishes as $\mu\to-2^{+}$ (see Fig.~\ref{fig: fig 3}B) and thus we can expand $\sigma$ in terms of a small parameter $c$ to find
\begin{equation}
\sigma = {\frac{\sqrt{\mu+2}}{2}}\pm i\left({\frac{\sqrt{2-\mu}}{2}} - \frac{c}{2\sqrt{4-\mu^2}}\right)+\mathcal{O}(c^2)\,.\label{eq: sigma}
\end{equation}
This expression shows that there are two different length scales impacting the front. First a decay length scale, given by $1/\Re \sigma\sim 1/\sqrt{\mu+2}$, which diverges as we approach the instability threshold. Second, an oscillation length scale, $1/\Im \sigma\sim 1/\sqrt{2-\mu}$, which stays of order $\mathcal{O}(1)$ and determines the non-monotonic behaviour---assuming that $c = o(\sqrt{2+\mu})$ near $\mu = -2$. The behaviour in the far field of the wave captures well the front dynamics (Fig.~\ref{fig: fig 3}E).

\subsection*{Strong cell-cell adhesion, invasion, and pattern formation}

When $\mu\leq -2$, the homogeneous steady state $\rho=1$ becomes unstable, and the system forms spatially localised cell aggregates (Fig.~\ref{fig: fig 4}A) behind a wave of invasion (Fig.~\ref{fig: fig 4}B). We first describe these aggregates in one spatial dimension before turning to two dimensions. In the strong-adhesion limit ($\mu\to -\infty$), the aggregate profile $\rho_s(\mathbf{x})$ can be obtained asymptotically by considering an expansion in the size of the aggregate. In one spatial dimension, the profile takes the form $\rho_s(x) = a_1[\cos(\sqrt{|\mu|}\, x) + 1]$, where the amplitude $a_1 = 2/3$ is determined by the integral constraint imposed by the logistic growth term (see Appendix~D.4). In particular, the peak density $\rho_{\max} = 2a_1 = 4/3$ exceeds the carrying capacity $\rho = 1$. Biologically, this reflects transient local overcrowding driven by adhesion-mediated aggregation: cells accumulate beyond the proliferative carrying capacity, and the excess is regulated by the logistic growth term, which acts as a restoring force.

Assuming radial symmetry in two dimensions, which is suggested by numerical simulations, $\rho_s(\mathbf{x})=\rho_s(r)$ with $r=|\mathbf{x}|$, the profile in two spatial dimensions approaches
\begin{align}
\frac{\rho_s(r)}{\rho_\mathrm{max}}
\sim
\frac{J_0\left(\sqrt{|\mu|}r\right)-J_0(j_{1,1})}{1-J_0(j_{1,1})},
\qquad 0\le r\le \ell,
\label{eq: steady state mu}
\end{align}
where $J_0$ is the Bessel function of the first kind, $\rho_{\max} \approx 1.74$, and the aggregate radius satisfies $\ell\sim j_{1,1}/\sqrt{|\mu|}$. Thus, the aggregates become smaller as adhesion strength increases.
This is the same spatial pattern found for cell aggregation without proliferation \cite{falco2022local}, but here the amplitude differs because it is modified by growth. We note that, although the approximate profile in Eq.~\eqref{eq: steady state mu} is derived in the limit $\mu \to -\infty$, it remains in excellent agreement with our numerical simulations even for values of $\mu$ close to the patterning threshold, $\mu=-2$ (Fig.~\ref{fig: fig 4}C). Moreover, from the linear stability analysis we obtain the characteristic wavelength of the pattern:
$\ell_a=2\pi\sqrt{{2}/{|\mu|}}$, 
 which sets the typical distance between spots---see Appendix~\ref{sec: aggregation}. Importantly, this regime supports cell invasion behind a patterned region of spots or cell aggregates, even in a single-species model, in contrast to classical Turing mechanisms that require at least two interacting species. The underlying mechanism is again the competition between processes acting at different length scales: adhesion is strong enough to drive aggregation, while diffusion and proliferation still sustain invasion.

\subsection*{Robust invasion from saturated adhesion}
In the unsaturated adhesion model, we found that strong adhesion can deeply alter invasion dynamics: planar fronts may lose monotonicity, develop transverse protrusions, or even break into aggregate-like patterns behind the invading edge. While such behaviours may be relevant in some contexts, they are unlikely to be desirable in developmental or tissue-repair settings where coordinated, cohesive spreading is required. This motivates us to consider a simple regulatory mechanism in which adhesion weakens as cells approach confluence.
To capture this, we introduce a density-dependent adhesion mobility $m(\rho) = \rho(1-\rho)$ \cite{CarrilloMurakawaCellAdhesion} so that adhesion is strongest at intermediate density and vanishes both at low density and near confluence. The model then reads
\begin{equation}
\partial_t\rho = \nabla\cdot\left(\rho(1-\rho)\nabla(\mu\rho-\Delta\rho)\right) + \alpha\nabla\cdot\left(\rho^2\,\nabla\rho \right) + \rho(1-\rho),
\label{eq: local model density-dep}\tag{$\mathrm{II}$}
\end{equation}
where, as in the unsaturated adhesion case, $\mu=\alpha-\omega\in\mathbb{R}$ and $\alpha>0$.
The key consequence of this choice is that the saturated state remains stable for all parameter values. Indeed, because $m(1)=0$, condition~\eqref{eq: B} is never satisfied, and therefore the saturated state $\rho=1$ does not undergo the adhesion-driven instability seen in the unsaturated adhesion model. Biologically, this means that the tissue can remain cohesive at high density without triggering the instabilities associated with uniformly strong adhesion.

For weak adhesion ($\alpha\to+\infty$ with $\omega=\mathcal{O}(1)$), the invasion dynamics are essentially unchanged from the unsaturated adhesion case: we recover fast monotonic travelling fronts. In fact, to leading order, the travelling-wave profile is the same as in Eq.~\eqref{eq: constant weak adhesion}, with the asymptotic wave speed $c\sim\sqrt{\mu/2}$---see Appendix~\ref{sec: weak adhesion saturated}.

\begin{figure}[t!]
    \centering
\includegraphics[width=\linewidth]{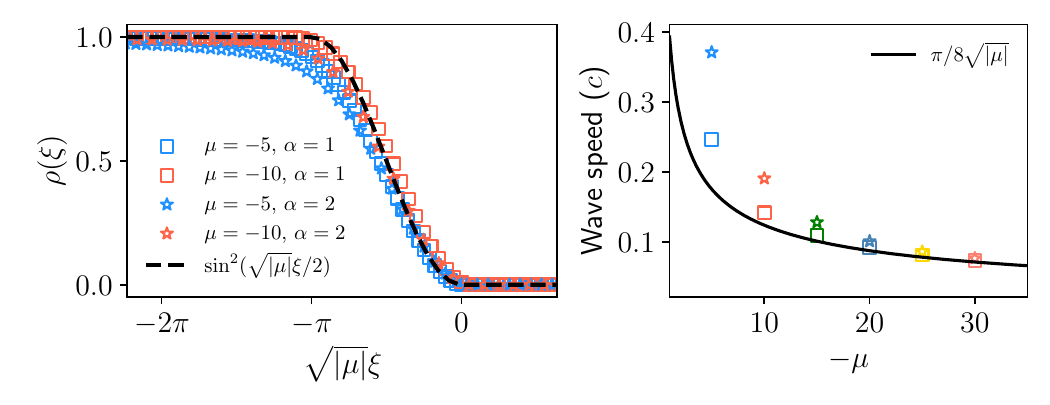}
    \caption{\textbf{Density-dependent adhesion gives rise to monotonic invasion fronts.}
(left) Convergence to the asymptotic solution in the limit $\mu \to -\infty$, observed even for relatively small $|\mu|$. 
(right) Asymptotic wave speeds and their convergence to the theoretical prediction. Squares and stars denote the cases $\alpha = 1$ and $\alpha = 2$, respectively.}
    \label{fig: fig 5}
\end{figure}

The difference becomes apparent in the strong-adhesion regime---see Appendix~\ref{sec: strong adhesion saturated}. When $\omega\to+\infty$ with $\alpha=\mathcal{O}(1)$, so that $\mu=\alpha-\omega\to-\infty$, the model still supports invasion, but now through slow monotonic fronts rather than oscillatory ones. This is because, although the adhesion parameter $\omega$ is large, the effective adhesive contribution near the saturated state vanishes: $m(1) = 0$ ensures that adhesion does not destabilise the tissue at high density, even though it remains strong at intermediate densities within the front. In one spatial dimension, and writing again $\xi=x-ct$, we obtain the asymptotic profile
\begin{align*}
\rho(\xi)\sim
\begin{cases} 1,&\mbox{ for }\quad -\infty < \xi \leq -\pi/\sqrt{|\mu|}\,;\ \\ \sin^2\left({\sqrt{|\mu|}\xi}/{2}\right) ,&\mbox{ for }\quad -\pi/\sqrt{|\mu|} < \xi \leq 0\,;\ \\ 0 \,,&\mbox{ for }\quad \xi > 0\,;\ \end{cases}
\end{align*}
with the asymptotic wave speed
$c\sim {\pi}/8{\sqrt{|\mu|}}$ (Fig.~\ref{fig: fig 5}). Hence, strong adhesion still slows tissue advance, but it no longer destabilises the invading front. Overall, this density-dependent mechanism preserves adhesion-mediated collective invasion while suppressing the front instabilities seen with unsaturated adhesion. In biological terms, it provides a simple route to robust tissue spreading: cells can maintain strong adhesive interactions at intermediate density, but relax them near confluence to avoid protrusions, stream breakup, or folding of the invading tissue.

\section*{Discussion}

Cell invasion and spatial pattern formation are widely regarded as distinct manifestations of cellular self-organisation. In this work, we have shown that both behaviours can emerge as different dynamical regimes of a single adhesion-driven mechanism. By analysing a local model of adhesion- and proliferation-driven self-organisation, we demonstrated that the interplay between short-range repulsion, longer-range adhesion, and density-dependent proliferation is sufficient to generate a broad spectrum of behaviours in one and two spatial dimensions, ranging from monotone invasion fronts to non-monotone travelling waves, localised aggregates, and fingering-like instabilities in two spatial dimensions. Furthermore, the relative analytical tractability of the model compared to non-local equivalents permits further analysis using asymptotic methods and travelling wave analysis, which allowed us to quantify the dynamics of migrating aggregates. 

In contrast to classical Turing theory, where pattern formation typically relies on at least two interacting species \cite{turing1952chemical,green2015positional}, robust spatial structure in this model can be produced in a one-species system through the competition between processes acting across distinct length scales. In this respect, the model sits alongside other non-classical self-organising mechanisms, such as mechanochemical patterning \cite{brinkmann2018post,recho2019theory}, while also connecting to recent formulations of directed cell migration \cite{yu2025directed} and to local and non-local continuum descriptions of adhesion-mediated collective behaviour \cite{buttenschon2024cells}. Biologically, our results show that cell--cell adhesion is not simply stabilising within collectives; in the unsaturated adhesion case (model (I)), increasing adhesion slows invasion and can destabilise the invasive front, while density-dependent regulation suppresses these instabilities and restores cohesive tissue expansion. In two spatial dimensions, the resulting protrusive fronts resemble generic fingering behaviours seen in invasive tissues and in physical descriptions of tissue spreading \cite{alert2019active}, and may be relevant to processes such as epithelial budding morphogenesis \cite{wang2021budding}, although a formal transverse stability theory is yet to be developed. This also motivates further study of the role of adhesive interactions in extracellular-matrix invasion settings, where cell--matrix adhesion is likewise important \cite{martinson2023dynamic, crossley2024modeling, kim2026collective}, and where the local model may provide an analytically tractable framework for further mechanistic understanding.

From a modelling perspective, the local formulation yields a robust theoretical framework while remaining well suited to numerical simulation \cite{falco2022local}. This tractability allowed us to derive explicit instability thresholds for the onset of non-monotonicity, asymptotic wave speeds and front profiles in both weak- and strong-adhesion limits, and approximate aggregate shapes and intrinsic pattern length scales in the strongly adhesive regime. The isotropic nature of the instability means that the model favours spot-like aggregates and localised fingering in two dimensions, suggesting that while the strength of adhesion may determine a local characteristic patterning length scale, stripe-like morphologies likely require additional ingredients, such as anisotropic interactions, responses to exogenous signals within the cellular microenvironment, or geometric confinement of the tissue domain. Overall, our results identify adhesion strength and its regulation by local density as key determinants of collective tissue dynamics, and open the door to further investigation of adhesion-mediated self-organisation in development, wound healing, and disease.

\section*{Acknowledgments}
The authors thank Ruth Baker, Giulia Celora, Dagmar Iber, and the other members of the Iber group for helpful discussions. For the purpose of Open Access, the authors have applied a CC BY public copyright licence to any Author Accepted Manuscript (AAM) version arising from this submission. CF acknowledges support from a Hooke Research Fellowship.

\section*{Data and Code Availability}

The code used to solve the model is available upon request and will be made publicly available upon publication. Numerical simulation relies on a finite volume scheme based on the method developed by Bailo et al.~\cite{bailo2021unconditional}---see Appendix~\ref{sec: finite volume}. All model and simulation parameters used to generate the main text figures can be found on Appendix~\ref{sec: parameters}.

\bibliography{refs}

\onecolumn
\appendix
\renewcommand\thefigure{\thesection.\arabic{figure}}    
\setcounter{figure}{0}    

\renewcommand{\theequation}{\thesection.\arabic{equation}}
\setcounter{equation}{0}    

\hspace{-.5cm}\large{\textbf{Supplementary Information}}
\vspace{1cm}
\normalsize
\tableofcontents
\pagenumbering{gobble}

\newpage
\pagenumbering{arabic}
\setcounter{page}{1}

\section{Short-range interactions approximation}
\label{sec:appendix-nonlocal-to-local}

Starting from Eq.~\eqref{eq:nonlocal_model} in the main text, we assume that adhesive interactions are short-ranged. We follow closely the derivation in \cite{falco2022local, falco2025nonlocal}. Note that such an approximation was already highlighted in the original model by Armstrong et al. \cite{ArmstrongPainterSherratt} and in \cite{murray2001mathematical} to describe long-range diffusion. The short-range approximation can be achieved, for instance, by rescaling the interaction kernel as
\[
    W(\mathbf{x}) = \varepsilon^{-d}\varphi(\mathbf{x}/\varepsilon),
\]
where $r = |\mathbf{x}|$ and $0<\varepsilon \ll 1$ denotes a small \emph{sensing radius}. We further assume that adhesive interactions are symmetric, so $\varphi(\mathbf{x})=\varphi(-\mathbf{x})$. With this rescaling, and omitting the time dependence in $\rho$, the spatial convolution can be written as
\begin{equation*}
    (W*\rho)(\mathbf{x})
    = \varepsilon^{-d}\int_{\Omega}\rho(\mathbf{x}-\mathbf{y})\,\varphi\!\left(\frac{\mathbf{y}}{\varepsilon}\right)\mathrm{d}\mathbf{y}
    = \int_{\Omega}\rho(\mathbf{x}-\varepsilon\mathbf{y})\,\varphi(\mathbf{y})\,\mathrm{d}\mathbf{y}.
\end{equation*}

The limit of short-range attraction can be obtained by expanding $\rho(\mathbf{x}-\varepsilon\mathbf{y})$ in a Taylor series around $\mathbf{x}$ for small values of the sensing radius $\varepsilon$:
\begin{align*}
    (W*\rho)(\mathbf{x})
    &= \rho(\mathbf{x})\int_{\Omega}\varphi(\mathbf{y})\,\mathrm{d}\mathbf{y}
    + \varepsilon\int_{\Omega}\!\left(\nabla\rho(\mathbf{x})\cdot\mathbf{y}\right)\varphi(\mathbf{y})\,\mathrm{d}\mathbf{y} \\
    &\quad + \frac{\varepsilon^2}{2}\int_{\Omega}\!\left(\mathbf{y}^T H_\rho(\mathbf{x})\mathbf{y}\right)\varphi(\mathbf{y})\,\mathrm{d}\mathbf{y}
    + o(\varepsilon^2),
\end{align*}
where $H_\rho(\mathbf{x})$ denotes the Hessian matrix of $\rho$.

We retain only the leading terms in this expansion, as they should provide a good approximation for short-range interactions.  
For the zeroth-order term, we have
\[
    \rho\int_{\Omega}\varphi(\mathbf{y})\,\mathrm{d}\mathbf{y}
    = M_0\,\rho,
\]
where $M_0$ denotes the zeroth moment of $\varphi$. The first-order term vanishes due to the symmetry of $\varphi$.  
Hence, the next non-vanishing contribution arises from the second-order term, which can be calculated as
\begin{align*}
    \int_{\Omega}
    \!\left(\mathbf{y}^T H_\rho(\mathbf{x})\mathbf{y}\right)\!
    \varphi(\mathbf{y})\,\mathrm{d}\mathbf{y}
    &= \sum_{i=1}^d\sum_{j=1}^d
    \frac{\partial^2\rho}{\partial x_i\partial x_j}
    \int_{\Omega}y_i y_j \varphi(\mathbf{y})\,\mathrm{d}\mathbf{y}, \\
    &= \sum_{i=1}^d
    \frac{\partial^2\rho}{\partial x_i^2}
    \int_{\Omega}y_i^2\varphi(\mathbf{y})\,\mathrm{d}\mathbf{y}, \\
    &= \frac{1}{d}
    \!\left(\int_{\Omega}|\mathbf{y}|^2\varphi(\mathbf{y})\,\mathrm{d}\mathbf{y}\right)
    \!\Delta\rho
    = \frac{M_2}{d}\,\Delta\rho,
\end{align*}
where $M_2$ is the second moment of $\varphi$. Here, again, we used the symmetry of $\varphi$.

Combining these results, we obtain
\begin{equation*}
    W*\rho
    = M_0\,\rho
    + \frac{\varepsilon^2 M_2}{2d}\,\Delta\rho
    +\mathcal{O}(\varepsilon^4)\,.
\end{equation*}
By using this approximation in the original non-local model we obtain the fourth-order equation
\begin{equation*}
    \partial_t\rho = \nabla\cdot\left[\left(D(\rho) -\Tilde{m}(\rho)\right)\nabla\rho - a\,\Tilde{m}(\rho)\nabla\Delta\rho\,\right] + r\rho(1 - \rho)\,,
\end{equation*}
where
\begin{align*}
    \Tilde{m}(\rho) = M_0 \,m(\rho)\,,\qquad a=\frac{\varepsilon^2 M_2}{2d M_0}\,.
\end{align*}

\section{Model without proliferation: gradient-flow structure}
\label{sec:appendix-energy}
By neglecting the proliferation term, the local model admits a gradient-flow structure with respect to a weighted metric. That is, ignoring logistic growth, we have
\begin{align*}
    \partial_t\rho & = \nabla\cdot\left[(d(\rho)-\omega m(\rho))\nabla\rho-m(\rho)\nabla\Delta\rho\,\right] 
    \\
    & = \nabla\cdot\left[m(\rho)\left(\left({d(\rho)}/{m(\rho)} - \omega \right)\nabla\rho - \nabla\Delta\rho\right)\right]
    \\
    & = \nabla\cdot\left[m(\rho)\left(\nabla h(\rho) - \nabla\Delta\rho\right)\right]\,,
\end{align*}
where $h(\rho)$ satisfies $h'(\rho) = d(\rho)/m(\rho) - \omega$. With this, the model can be written in the following form, which is useful for the choice of numerical scheme:
\begin{equation}
    \partial_t\rho = \nabla\cdot\left[m(\rho)\nabla\frac{\delta\mathcal{F}}{\delta\rho}\right], \label{eq: weighted gradient flow}
\end{equation}
with the free energy
\begin{equation*}
    \mathcal{F}[\rho] = \frac{1}{2}\int_\Omega |\nabla\rho|^2\,\mathrm{d}\mathbf{x} + \int_\Omega H(\rho)\,\mathrm{d}\mathbf{x}\,
\end{equation*}
where $H' = h$. The function $H$ is determined up to addition of any linear function, which does not change the convexity/concavity of $H$. We note that Eq.~\eqref{eq: weighted gradient flow} is of the form of a weighted 2-Wasserstein gradient flow \cite{antonio2024competing}, so the free energy is non-increasing in time, i.e.
\begin{equation*}
    \frac{\mathrm{d}}{\mathrm{d}t}\mathcal{F}[\rho] = - \int_\Omega m(\rho)\left|\nabla\frac{\delta\mathcal{F}}{\delta\rho}\right|^2\,\mathrm{d}\mathbf{x}\leq 0\,.
\end{equation*}

In particular, energy minima of Eq.~\eqref{eq: weighted gradient flow} will satisfy
\begin{equation*}
    \frac{\delta\mathcal{F}}{\delta\rho} = -\Delta\rho + h(\rho) = C\,,
\end{equation*}
for a constant $C$. For the unsaturated adhesion model $(\mathrm{I})$, we have $h(\rho) = \mu\rho$ and hence pattern formation is only possible when $\mu < 0$---adhesion is sufficiently strong. In this case, the model predicts the formation of spots or aggregates with an explicit shape, calculated in \cite{falco2022local}, which in two-spatial dimensions reads\footnote{Note the different definition of $\mu$ here compared to that in \cite{falco2022local}.}
\begin{equation*}
\rho (\mathbf{x}) = \frac{N_\rho|\mu|}{\pi j_{1,1}^2}\left(1-\frac{J_0(\sqrt{|\mu|} \textbf{x}|)}{J_0(j_{1,1})}\right), \quad |\textbf{x}| \leq \frac{j_{1,1}}{\sqrt{|\mu|}}\, ,
\end{equation*}
where $J_0$ is the Bessel function of the first kind, $j_{1,1}$ is the first zero of $J_1$, and $N_\rho$ is the number of cells in the cell aggregate.

For the general model, linear stability analysis around the homogeneous steady state $\rho = \phi$ reveals that perturbations of the form $\psi(\mathbf{x},t) = e^{\lambda t+i\mathbf{k}\cdot\mathbf{x}}$ will grow with rate
\begin{equation*}
    \lambda(k;\phi) = -m(\phi)|\mathbf{k}|^2\left(|\mathbf{k}|^2 + H''(\phi)\right)\,.
\end{equation*}
In particular, pattern formation occurs when adhesion is sufficiently strong at $\rho = \phi$, this is: \begin{equation*}
    H''(\phi) < 0\iff \omega > \frac{d(\phi)}{m(\phi)}\,.
\end{equation*}
\section{Conditions for pattern formation}
\label{sec:linear-stability}
We analyse the stability of homogeneous steady states of the full model with proliferation, Eq.~\eqref{eq: local model constant}, under small perturbations.  
The reaction term $\rho(1-\rho)$ admits two homogeneous steady states, $\phi = 0$ and $\phi = 1$.  
We thus consider perturbations of the form $\rho(\mathbf{x},t) = \phi + \psi(\mathbf{x},t)$
and linearise the resulting equation to obtain 
\begin{equation*}
    \partial_t \psi
    = \left(d(\phi) - \omega m(\phi)\right)\Delta\psi
    - m(\phi)\Delta^2\psi
    + (1 - 2\phi)\psi\,.
\end{equation*}
We seek solutions of the form
$\psi(\mathbf{x},t) = e^{\lambda t + i\mathbf{k}\cdot\mathbf{x}}$, with $k = |\mathbf{k}|$, which yields the dispersion relation
\begin{equation*}
    \lambda(k;\phi) = (1 - 2\phi)
    - \left(d(\phi) - \omega m(\phi)\right)k^2
    - m(\phi)k^4\,.
\end{equation*}
From here, we see that $\phi = 0$ is linearly unstable as
\begin{equation*}
    \lambda(k;0) = 1-d(0)k^2>0\,,\quad\mbox{if } \quad k^2 < \frac{1}{d(0)}\,.
\end{equation*}
This condition will be satisfied for a large enough domain.

On the other hand, for $\phi = 1$, we obtain
\begin{equation}
    \lambda(k;1) = -1 -(d(1)-\omega m(1))k^2-m(1)k^4\,.\label{eq: lambda la}
\end{equation}
which yields that $\phi = 1$ is linearly stable if $d(1) \geq \omega m(1)$---which includes the case $m(1) = 0$. When $d(1) < \omega m(1)$, the growth rate $\lambda(k;1)$ has a maximum when $k^2 = [\omega - d(1)/m(1)]/2$, which yields
\begin{equation*}
    \lambda_\mathrm{max} = -1 + \frac{(d(1)-\omega m(1))^2}{4m(1)}\,,
    %\label{eq: max wavelength}
\end{equation*}
so $\phi = 1$ can become unstable when $(d(1)-\omega m(1))^2 > 4m(1)$. Summarising, for pattern formation, we need both homogeneous steady states to be unstable, and hence we obtain the necessary conditions
\begin{align}
       (\phi = 1 \quad \text{unstable}) \quad &  \omega m(1) -d(1)  > 2\sqrt{m(1)}\,.\tag{$\dagger$} \label{eq: B sup}
\end{align}

\section{Unsaturated adhesion model}
\label{sec: constant appendix}
In this appendix we analyse the unsaturated adhesion model given by Eq.~\eqref{eq: local model constant}.
\subsection{Travelling fronts}
\label{sec: travelling fronts}
We look for travelling fronts with speed $c = c(\mu) >0$ in the horizontal direction in the unsaturated adhesion model given by Eq.~\eqref{eq: local model constant}. We thus set $\rho(\mathbf{x},t) = \mathcal{P}(\xi)$ for $\xi = x-ct$, which yields\footnote{This definition is valid up to a constant which fixed the translational invariance to $\mathcal{P}(0) = 0$.}
\begin{equation*}
    c\mathcal{P}' - (\mathcal{P}\mathcal{P}''')' +\mu (\mathcal{P}\mathcal{P}')' + \mathcal{P}(1-\mathcal{P}) = 0\,,\tag{C.1} \label{eq: tw fronts sup}\end{equation*}
satisfying $\mathcal{P}(-\infty) = 1, \,\mathcal{P}(0) = 0$. We note that the first condition sets two degrees of freedom as the unstable manifold at $\mathcal{P} = 1$ has dimension 2---see Eq.~\eqref{eq: ch poly} and discussion below. There are two further boundary conditions at $\mathcal{P}=0$, which can be understood from a local analysis of the time-dependent problem near $\rho = 0$. We return briefly to the problem  given by Eq.~\eqref{eq: local model constant}, and consider, in one spatial dimension, $\rho = 0$ at $x = s(t)$. Next, for $z<0$, and $0<\varepsilon\ll1$ we consider $x = s(t) + \varepsilon z$, and rescale the density accordingly, $\rho = \varepsilon^3 \tilde\rho$. In terms of time, the rescaled spatial variable $z$, and the rescaled density $\tilde\rho$, we obtain
\begin{equation*}
    \varepsilon \partial_t\tilde\rho = -\partial_z\left[\tilde\rho\left(\partial_z^3\tilde\rho - \dot{s}-\varepsilon^2\mu\partial_z\tilde\rho\right)\right] + \varepsilon\tilde\rho(1-\varepsilon^3\tilde\rho)\,.
\end{equation*}
At leading order in $\varepsilon$, we obtain
\begin{equation*}
    \partial_z\left[\tilde\rho\left(\partial_z^3\tilde\rho-\dot{s}\right)\right] = 0\,.
\end{equation*}
Using that $\tilde\rho = 0$ at $z = 0$, we obtain $\dot{s}=\partial_z^3\tilde\rho$. For a travelling wave solution, in particular, we obtain $\dot{s} = c$ and $\mathcal{P}'''(0) = c$, which gives one additional boundary condition.

The last boundary condition at $\mathcal{P} = 0$ can be understood from a complementary local analysis. Substituting the ansatz $\tilde{\rho}\sim A(t)z$ into Eq.~\eqref{eq: local model constant}, and performing a local analysis near the interface, we obtain the asymptotic expansion
\begin{equation*}
    \tilde{\rho}\sim A(t) z + B(t)z^2 + \left(\frac{\mu A(t) - \dot{s}}{6}\right)z^3 + o(z^3)\,.
\end{equation*}
Thus, consistency with the boundary condition at $z = 0$ given by $\dot{s}=\partial_z^3\tilde\rho$, also requires $A = 0$\footnote{Here we assumed $\mu\neq 0$. In the particular case where $\mu = 0$ it may be possible to find solutions with a non-zero contact angle.}. This is, at $z = 0$ we have the additional condition $\partial_z\tilde\rho = 0$, which in the travelling wave frame reads: $\mathcal{P}'(0) = 0$. As a result, we look for solutions to Eq.~\eqref{eq: tw fronts sup} together with the boundary conditions
\begin{equation*}
    \mathcal{P}(-\infty) = 1, \quad  \mathcal{P}(0) = \mathcal{P}'(0) = 0\,,\quad \mathcal{P}'''(0) = c\,.
\end{equation*}

Returning to the travelling wave problem, we further note that $\mathcal{P} = 0$ is always unstable. The stability of the state $\mathcal{P} = 1$ depends on $\mu$, with the critical necessary condition for the existence of travelling front given by $\mu \geq -2$ 
--- condition~\eqref{eq: B sup}.

We can rewrite Eq.~\eqref{eq: tw fronts sup}  as a four-dimensional dynamical system, denoting $\mathrm{p}(\xi) = \mathcal{P}$, $\mathrm{q} (\xi)= \mathcal{P}'$, $\mathrm{r}(\xi) = \mathcal{P}''$,   $\mathrm{s}(\xi) = \mathcal{P}'''$, as 
\begin{equation*}
    \begin{cases}
       \mathrm{p}' & = \mathrm{q}\,,
        \\
       \mathrm{q'} & =\mathrm{r}\,,
       \\
      \mathrm{r}' & = \mathrm{s}\,,
       \\
        \mathrm{p} \mathrm{s'} & = c \mathrm{q}  +\mu\mathrm{q}^2-\mathrm{q}\mathrm{s}+\mu \mathrm{p}\mathrm{r} + \mathrm{p}(1 - \mathrm{p})\,.
    \end{cases}
\end{equation*}
From here, we obtain that the steady state at the back of the wave corresponds to $\mathrm{p} = 1$, giving $(\mathrm{p},\mathrm{q},\mathrm{r},\mathrm{s}) = (1,0,0,0)$, but the system is singular at $\mathrm{p}= 0$ and can be desingularised by using the change of variables $\mathrm{p\,}\mathrm{d}/\mathrm{d}\xi = \mathrm{d}/\mathrm{d}\mathrm{z}$, which gives
\begin{equation}
    \begin{cases}
       \mathrm{p}' & =\mathrm{p} \mathrm{q}\,,
        \\
       \mathrm{q'} & =\mathrm{p}\mathrm{r}\,,
       \\
      \mathrm{r}' & = \mathrm{p}\mathrm{s}\,,
       \\
         \mathrm{s'} & = c \mathrm{q}  +\mu\mathrm{q}^2 -\mathrm{q}\mathrm{s}+\mu\mathrm{p}\mathrm{r} + \mathrm{p}(1 - \mathrm{p})\,. 
    \end{cases}\label{eq: SI des odes}
\end{equation}
now with steady states $(\mathrm{p},\mathrm{q},\mathrm{r},\mathrm{s}) = (1,0,0,0)$, $(0,(\mathrm{s}_0-c)/\mu ,\mathrm{r}_0, \mathrm{s}_0)$, and $(0,0,\Tilde{\mathrm{r}}_0, \Tilde{\mathrm{s}}_0)$ for any real values of $\mathrm{r}_0, \mathrm{s}_0, \Tilde{\mathrm{r}}_0, \Tilde{\mathrm{s}}_0$. Note that if $\mathrm{s}_0 = c$ then the two possible steady states at $\mathcal{P} = 0$ have a zero contact angle, which is in agreement with the asymptotic result from the local analysis.  

The Jacobian of the system at $(1,0,0,0)$ reads
\[
J_1= 
\begin{pmatrix}
0 & 1 & 0 & 0 \\[6pt]
0 & 0 & 1 & 0 \\[6pt]
0 & 0 & 0 & 1 \\[6pt]
-1 & c & \mu & 0
\end{pmatrix},
\]
with characteristic polynomial
\begin{equation}
   q_{\mu,c}(\lambda) =   \lambda^4 -\mu\lambda^2-c\lambda+1\,.
   \label{eq: ch poly}
\end{equation}
The roots of this polynomial can have different behaviours depending on the values of $\mu$ and $c$. In particular, the roots with $\Re\lambda > 0$ define the unstable manifold of our system, which always has dimension two, and will approximate the travelling front near $\mathcal{P}=1$. Oscillatory behaviour and hence, non-monotonic waves, are expected for roots with $\Im \lambda \neq 0$. The critical threshold for the existence of non-monotonic fronts occurs when $q_{\mu,c}$ has a double real root, in which case, $\lambda$ satisfies
\begin{equation*}
    q_{\mu,c}(\lambda) = 0\,,\quad q_{\mu,c}'(\lambda) = 4\lambda^3-2\mu\lambda-c = 0\,.
\end{equation*}
This threshold can be described by the parametric curve $s\mapsto(\tilde{c},\tilde\mu)$ given by:
\begin{align}
    \tilde\mu(s) &= 3s^2 - \frac{1}{s^2}\,, \label{eq: curve 1}
    \\
    \tilde c(s) &= -2s^3+\frac{2}{s}\,, \label{eq: curve 2}
\end{align}
for $s\in[1/\sqrt 3,1]$, which yields $\tilde c(s)\in [0,16\sqrt 3/9]$, and $\tilde\mu(s)\in [-2,2]$. This implies that roots are real for $\mu >2$, such that fronts are expected to be monotonic and can be complex, such that oscillatory behaviour is expected for $\mu <2$. The transition from monotonic waves to non-monotonic waves will occur when the actual wave speed $c(\mu)$ intersects the curve given by Eqs.~\eqref{eq: curve 1} and~\eqref{eq: curve 2}, at $\mu = \mu_c$. We can estimate this critical threshold $\mu_c$ by using the asymptotic behaviour for the wave speed $c(\mu)$ for $\mu\gg 1$. In this regime---this will be justified rigorously below---we expect the second-order diffusion to dominate over the fourth-order term, obtaining the Porous-Fisher model in the limit $\mu\rightarrow + \infty$. The Porous-Fisher model  in one spatial dimension,
\begin{equation*}
    \partial_t\rho = \mu\partial_x(\rho\,\partial_x\rho) + \rho(1-\rho)\,,
\end{equation*}is known to give rise to travelling wave solutions with compact support and with the explicit wave speed $c=\sqrt{\mu/2}$ \cite{murray2001mathematical}. Using this approximation, and substituting this wave speed value in Eqs.~\eqref{eq: curve 1}, and~\eqref{eq: curve 2}, we obtain the following approximation for the non-monotonicity threshold:
\begin{equation*}
    \mu_c \approx 1.18\,.
\end{equation*}

\subsection{Weak cell-cell adhesion}\label{sec: weak adhesion constant} When adhesion is weak ($\mu \gg 1$), we expect the travelling front to approach the solution of the Porous-Fisher equation. Formally, under a dilation of space by a factor of $\sqrt{\mu}$, our model converges to the well-known Porous-Fisher model. This motivates seeking asymptotic solutions to travelling fronts on two spatial scales, giving rise to an \emph{outer} (large $\xi$) and an \emph{inner} (small $\xi$) solution. 

\underline{\emph{Outer solution I} ($\mathcal{P}\sim\mathcal{P}_{\mathrm{out}}$ for $\xi=\mathcal{O}(\sqrt{\mu})$).} Anticipating that the speed tends to that of the Porous-Fisher model, we rescale $c = \sqrt{\mu}\nu$. A dominant balance then occurs for $\xi = \sqrt{\mu}\eta$ and yields (keeping primes to denote differentiation with respect to $\eta$):
\begin{equation}
    \nu \mathcal{P}' - \mu^{-2}(\mathcal{P}\mathcal{P}''')' + (\mathcal{P}\mathcal{P}')' + \mathcal{P}(1-\mathcal{P}) = 0\,.\label{eq: outer}
\end{equation}
At leading order (formally $\mu = +\infty$), we obtain the Porous-Fisher equation in travelling wave coordinates,
\begin{equation*}
    \nu_0\mathcal{P}_0'+(\mathcal{P}_0\mathcal{P}_0')' + \mathcal{P}_0 (1-\mathcal{P}_0) = 0\,, \tag{outer $\mathcal{O}(1)$}
\end{equation*}
with the well-known solution \cite{murray2001mathematical}
\begin{align*}
    \mathcal{P}_0(\eta) & = 1 - e^{\eta/\sqrt{2}}\,,\quad\mbox{for}\quad \eta \leq 0\,;
    \\
    \nu_0 & = \frac{1}{\sqrt{2}}\,.
\end{align*}
Equivalently, in terms of $\xi$ we obtain $\mathcal{P}(\xi)\sim \mathcal{P}_{0}(\xi)= 1 -e^{\xi/\sqrt{2\mu}}$, as $\mu\rightarrow +\infty$. Solutions with $\nu_0 > 1/\sqrt{2}$ are also possible \cite{murray2001mathematical}, but these have infinite tails, which we do not observe in our simulations. A phase plane argument yields the selected speed.

Next we look for a perturbative expansion
\begin{align*}
    \mathcal{P}_\mathrm{out} & = \mathcal{P}_0 + \mu^{-1}\mathcal{P}_1 + \mu^{-2}\mathcal{P}_2 + o(\mu^{-2})\,,
    \\
    \nu & = \nu_0 +\mu^{-1}\nu_1 +\mu^{-2}\nu_2 + o(\mu^{-2})\,,
\end{align*}
where the small parameter in the expansion is determined by the inner solution (see below).

At order $O(\mu^{-1})$ we find
\begin{equation}
\left[\mathcal{P}_0\left(\mathcal{P}_1'+\frac{1}{\sqrt{2}}\mathcal{P}_1\right)\right]' + (1-2\mathcal{P}_0)\mathcal{P}_1 = \frac{\nu_1}{\sqrt{2}}(1-\mathcal{P}_0)\,. \tag{outer $\mathcal{O}(\mu^{-1})$}
\end{equation}
It is helpful to rewrite the equation in terms of $p = \mathcal{P}_0(\eta)$, so that $\mathrm{d}/\mathrm{d}\eta = - \nu_0(1-p)(\mathrm{d}/\mathrm{d}p)$. The resulting equation can be written in compact form as 
\begin{equation*}
    \frac{1}{2}\frac{\mathrm{d}}{\mathrm{d}p}\left[p\frac{\mathrm{d}}{\mathrm{d}p}\left((1-p)\mathcal{P}_1\right)\right] + \frac{1-2p}{1-p}\mathcal{P}_1 = \frac{\nu_1}{\sqrt{2}}\,.
\end{equation*}
Next we note that at $p = 1$, $\mathcal{P}_1(1) = 0$. In particular, for a valid asymptotic series, we require that $\mathcal{P}_1$ approaches 0 not slower than the leading order solution $\mathcal{P}_0(p) = 1 -p$, so we expect $\mathcal{P}_1(p) =\mathcal{O}((1-p)^a)$ for $a> 1$ as $p\rightarrow 1$. This regularity requirement allows us to deduce $\nu_1$ without solving the full ODE exactly:
\begin{equation*}
    \frac{\nu_1}{\sqrt{2}} = \lim_{p\rightarrow 1}\left( -\frac{1}{2}(a+1)(1-p)^a+\frac{1}{2}a(a+1)p(1-p)^{a-1} +(1-2p)(1-p)^{a-1}\right) = 0\,.
\end{equation*}
Since $\nu_{1}=0$, the differential equation simplifies and can be solved explicitly to give
\begin{equation*}
\mathcal{P}_1(p) = A_1(1-p)\,,
\end{equation*}
where $A_1$ is a constant to be determined by matching with the inner solution. We note that the general solution also includes a second linearly independent solution that does not tend to zero as $p \to 1$; this solution is discarded by the regularity requirement above. In terms of the travelling wave coordinate, and for $\xi =\mathcal{O}(1)$,  we have so far
\begin{equation}
    \mathcal{P}_\mathrm{out}(\xi)=-\frac{\xi}{\sqrt{2\mu}} + \frac{4A_1 - \xi^2}{4\mu} + o(\mu^{-1})\,. %\label{eq: outer expand O(1)}
    \label{labelForCorrection}
\end{equation}

In fact, the correction to the wave speed arises at $O(\mu^{-2})$. To obtain this, we must first consider the inner region, which also determines the constant $A_{1}$.

\underline{\emph{Inner solution} ($\mathcal{P}\sim\mathcal{P}_{\mathrm{in}}$ for $\xi=\mathcal{O}(1/\sqrt{\mu})$).} The outer solution is no longer valid when $\xi$ approaches 0, since $\mathcal{P}_\mathrm{out}$ satisfies $\mathcal{P}(0) = 0$, but not the zero contact angle condition $\mathcal{P}'(0) = 0$. In this region, a dominant balance can be achieved by rescaling $\xi = \zeta/\sqrt{\mu}$, and $\mathcal{P} = \Tilde{\mathcal{P}}/\mu$, which yields
\begin{equation*}
\nu\Tilde{\mathcal{P}}' - (\Tilde{\mathcal{P}}\Tilde{\mathcal{P}}''')' + (\Tilde{\mathcal{P}}\Tilde{\mathcal{P}}')' + \mu^{-1}\Tilde{\mathcal{P}} - \mu^{-2}\Tilde{\mathcal{P}}^2=0\,.
\end{equation*}
The leading order problem is thus
\begin{equation*}
\nu_0\Tilde{\mathcal{P}}'_0 - (\Tilde{\mathcal{P}}_0\Tilde{\mathcal{P}}_0''')' + (\Tilde{\mathcal{P}_0}\Tilde{\mathcal{P}}'_0)' =0\,,\tag{inner
 $\mathcal{O}(1)$}
\end{equation*}
with the explicit solution
\begin{equation*}
   \tilde{\mathcal{P}}_0(\zeta) = \nu_0\left(e^\zeta - \zeta - 1\right)\,.
\end{equation*}
In terms of $\xi$ we obtain $\mathcal{P}(\xi)\sim \mathcal{P}_{\mathrm{in}}(\xi)\sim \nu_0(e^{\sqrt{\mu}\xi}-\sqrt{\mu}\xi - 1)/\mu$. In particular, for $\xi =\mathcal{O}(1)$ and large $\mu$ we obtain
\begin{equation*}
    \mathcal{P}_\mathrm{in}(\xi)= -\frac{\xi}{\sqrt{2\mu}} +\mathcal{O}(\mu^{-1})\,,
\end{equation*}
which agrees with the outer solution up to leading order ($O(\mu^{-1/2})$) when $\xi =\mathcal{O}(1)$.

The next-order problem is
\begin{equation*} \left[\tilde{\mathcal{P}}_0(\tilde{\mathcal{P}}_1'''-\tilde{\mathcal{P}}_1')\right]' = \tilde{\mathcal{P}}_0\,.\tag{inner
 $\mathcal{O}(\mu^{-1})$}
\end{equation*}
We do not need to solve this exactly to obtain the information required for matching. Instead, we determine its asymptotic behaviour as $\zeta \to \infty$. Integrating once, for $-\zeta \gg 1$, we find:
\begin{equation*}
        \tilde{\mathcal{P}}_1''' - \tilde{\mathcal{P}}
_1' = \frac{\zeta}{2}+\frac{1}{2}+\mathcal{O}(\zeta^{-1})\,.
\end{equation*}
Thus:
\begin{equation*}
    \tilde{\mathcal{P}}_1(\zeta) \sim -\frac{\zeta^2}{4}-\frac{\zeta}{2}+\mathcal{O}(\log\zeta),\quad\mbox{ as }\zeta\rightarrow -\infty\,.
\end{equation*}
Writing the first two terms of the inner expansion as $\zeta \to -\infty$ in terms of the outer variable $\xi$ for matching purposes, we find
\begin{equation}
      \mathcal{P}_\mathrm{in}(\xi)\sim -\frac{\xi}{\sqrt{2\mu}} - \left(\frac{1}{\sqrt{2}}+\frac{\xi^2}{4}\right)\frac{1}{\mu}+o(\mu^{-1})\,,
\end{equation}
which agrees with the outer solution (Eq.~\eqref{labelForCorrection}) if we set $A_1 = -1/\sqrt{2}$.

\underline{\emph{Outer solution II: Correction to the wave speed. }} At order $O(\mu^{-2})$ we find a non-zero correction to the wave speed. We use the outer expansion, and at this order we find  
\begin{equation*}
\left[\mathcal{P}_0\left(\mathcal{P}_2'+\nu_0\mathcal{P}_2\right)\right]' + (1-2\mathcal{P}_0)\mathcal{P}_2 = \nu_0\left(\nu_2-\nu_0^3+\nu_0(1-\mathcal{P}_0)\right) \,.\tag{outer
 $\mathcal{O}(\mu^{-2})$}
\end{equation*}
Again, we rewrite the equation in terms of $p = \mathcal{P}_0(\eta)$. The resulting equation can be written in compact form as 
\begin{equation*}
    \frac{1}{2}\frac{\mathrm{d}}{\mathrm{d}p}\left[p\frac{\mathrm{d}}{\mathrm{d}p}\left((1-p)\mathcal{P}_2\right)\right] + \frac{1-2p}{1-p}\mathcal{P}_2 = \nu_0\left(\nu_2-\nu_0^3+\nu_0(1-p)\right) \,.
\end{equation*}
The same regularity argument near $p = 1$ that we used to determine $\nu_1$ can applied here, giving
\begin{equation*}
    \nu_2 = \nu_0^3 = \frac{1}{2\sqrt{2}}\,.
\end{equation*}

We therefore obtain:
\begin{equation*}
    c = \sqrt{\frac{\mu}{2}} + \frac{1}{2\sqrt{2}\mu^{3/2}} + o(\mu^{-3/2})\,.
\end{equation*}

\subsection{Explicit travelling wave solutions with non-zero contact angle}\label{sec: finite contact angle} 
\begin{figure}[h!]
    \centering
\includegraphics[width=0.6\linewidth]{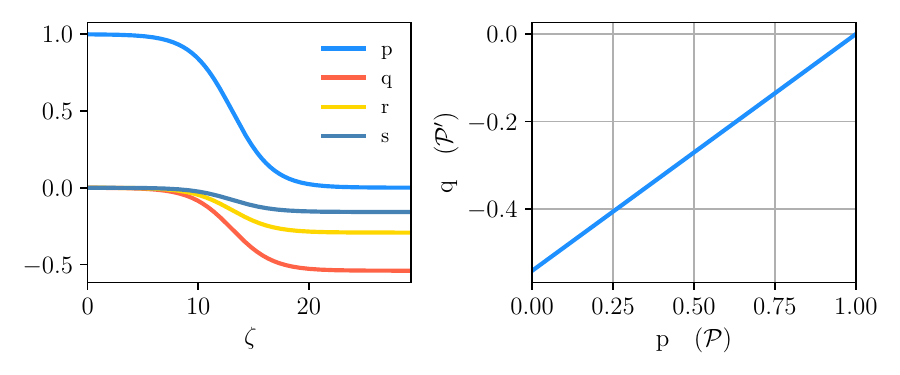}
    \caption{\textbf{Non-zero contact angle travelling waves. } Numerical solution of the system of ordinary differential equations given by Eqs.~\eqref{eq: SI des odes} with $\mu = 2$ and $c$ given by Eq.~\eqref{eq: c finite contact angle SI} and with initial condition at $\zeta = 0$: $(\mathrm{p}, \mathrm{q}, \mathrm{r}, \mathrm{s}) = (1,0,0,0)$ plus a small perturbation in the direction of the unstable manifold. Left panel shows the evolution in the desingularised travelling wave coordinate. Right panel shows a linear relationship between $\mathcal{P}'$ and $\mathcal{P}$ (Eq.~\eqref{eq: finite contact angle SI}), reminiscent of Porous-Fisher travelling waves \cite{murray2001mathematical}.}
    \label{fig: odes SI}
\end{figure}

The structure of the outer solution in the previous section also motivates looking for monotonically decreasing explicit solutions of the form $\mathcal{P}(\xi) = 1 -e^{\sigma \xi}$. Although these solutions have not been observed in time-dependent simulations of the PDE---and would be inconsistent with the local analysis of the interface in the time-dependent problem---, they are indeed solutions in the comoving frame, as verified analytically and numerically (Fig.~\ref{fig: odes SI}). We derive them here for completeness. Note that this type of profile approaches $\mathcal{P}=0$ with a finite contact angle ($\mathcal{P}'(0)\neq 0$). Upon substitution of this ansatz to Eq.~\eqref{eq: tw fronts sup}, we obtain two polynomial conditions
\begin{align*}
    q_{\mu,c}(\sigma) & = 0\,,
    \\
    2\sigma^4 - 2\mu\sigma^2 + 1& = 0\,,
\end{align*}
where the first equation says that $\sigma$ is a root of the characteristic polynomial of the system near $\mathcal{P} = 1$, and the second condition comes from the nonlinear terms in the equation. From here, we can choose a pair $(\sigma,c)$ satisfying both conditions, which yields two explicit and exact travelling wave solutions, given by
\begin{align*}
c & = \frac{1}{2\sigma}\,,
\\
   \sigma & =  \sigma_{\pm} = \sqrt{\frac{\mu\pm\sqrt{\mu^2-2}}{2}}\,.
\end{align*}
We note that these expressions are valid as long as $\mu > \sqrt{2}$. Hence we obtain the exact travelling wave solutions
\begin{align}
    \mathcal{P}(\xi) & = 1 - e^{\sigma_\pm\xi}\,,\quad\mbox{for}\quad\xi\leq 0\,;\label{eq: finite contact angle SI}
    \\
    c(\mu) &= \frac{1}{2}\sqrt{\frac{2}{\mu\pm\sqrt{\mu^2-2}}}\,,\quad\mbox{for}\quad\mu>\sqrt{2}\,.\label{eq: c finite contact angle SI}
\end{align}

\subsection{Cell invasion with aggregation}
\label{sec: aggregation}
Travelling fronts satisfy $\rho\rightarrow 1$ at the back of the wave. For strong cell-cell adhesion, such travelling fronts become unstable and cell aggregates start to form. This occurs when $\mu < -2$. We look for the shape profile, $\rho_s = \rho_s(\mathbf{x})$, of these cell aggregates, which satisfy the steady state equation
\begin{equation}
    \label{eq: steady state}
\nabla\cdot\left[\rho_s\nabla\,\left(\Delta\rho_s + |\mu|\rho_s\right)\right] = \rho_s(1-\rho_s)\,,
\end{equation}
which in general does not admit explicit solutions in terms of elementary functions. In particular, we look for symmetric, radially decreasing, compactly supported steady states around $\mathbf{x} = 0$, that is $\rho_s(\mathbf{x}) = \rho_s(|\mathbf{x}|)$ with $\rho_s = 0$ for $|\mathbf{x}| = \ell$, for an aggregate of radius $\ell$. Using the divergence theorem, and that steady states are compactly supported, we have
\begin{align}
    \int_{|\mathbf{x}| < \ell} \rho_s(1-\rho_s)\,\mathrm{d}\mathbf{x} & =   \int_{|\mathbf{x}| < \ell}\left[ \nabla\cdot(\rho_s\nabla\Delta\rho_s) + |\mu|  \nabla\cdot (\rho_s\nabla\rho_s)\right] \,\mathrm{d}\mathbf{x} = 0\,.\label{eq: condition 
    2d ss}
   % \\
   % & = \int_{|\mathbf{x}| = \ell}\rho_s_s \left[ \nabla\Delta\rho_s_s - \mu  \nabla\rho_s_s)\right] \cdot \mathbf{n}\, \mathrm{d}
\end{align}
Given that $\rho_s$ is non-negative, this requires $\rho_s(0)  > 1$. In one spatial dimension we would find
\begin{equation}
    \int_{-\ell}^\ell \rho_s (1-\rho_s)\,\mathrm{d}x = 0\,. \label{eq: condition 1d ss}
\end{equation}

To make analytical progress, we exploit the fact that $|\mu| > 2$, and perform matched asymptotics using $|\mu|^{-2}$ as a small parameter. It is particularly informative to present the approximate solution in one spatial dimension first.

\underline{\emph{Steady states in 1D}. }
In one spatial dimension, Eq.~\eqref{eq: steady state} reads
\begin{equation*}
\left[\rho_s\left(\rho_s''+|\mu|\rho_s\right)'\right]' = \rho_s(1-\rho_s)\,,
\end{equation*}
supplemented with boundary conditions $\rho_s(\ell)=\rho_s'(\ell)=0$, where the support length $\ell>0$ is unknown. Primes denote differentiation with respect to the spatial variable $x$. We assume that steady states are symmetric about the origin, which by translation invariance, we can set to $x=0$.

For strong adhesion ($-\mu\gg1$), we expect the steady-state cell aggregate to be localised on a small spatial scale of order $|\mu|^{-1/2}$. We therefore introduce the rescaled spatial variable $x={\hat{x}}/{\sqrt{|\mu|}}$.
In terms of $\hat{x}$, the steady state equation becomes
\begin{equation*}
\left[\rho\left(\rho''+\rho\right)'\right]' = |\mu|^{-2}\rho(1-\rho)\,,
\end{equation*}
where primes now denote differentiation with respect to $\hat{x}$. At leading order, we thus obtain
\begin{equation*}
\left[\rho\left(\rho''+\rho\right)'\right]' = 0\,.
\end{equation*}
Imposing symmetry about $\hat{x}=0$ yields the leading order solution
\begin{equation*}
\rho(\hat{x}) = a_1\cos\hat{x}+a_2\,,
\end{equation*}
with constants $a_1,a_2$. Requiring compact support and non-negativity gives $a_2=a_1$, so that, in the original spatial variable,
\begin{equation*}
\rho_s(x) = a_1\left[\cos\left(\sqrt{|\mu|}x\right)+1\right]
+ \mathcal{O}(|\mu|^{-2}),
\qquad |x|\le \ell,
\end{equation*}
where the half-length of the steady state is
\begin{equation*}
\ell=\frac{\pi}{\sqrt{|\mu|}}\,.
\end{equation*}
This leading-order steady state coincides with the non-proliferative result derived in \cite{falco2022local}.

Near the boundary $x=\pm\ell$, proliferation effects generate a narrow boundary layer whose contribution to $\rho_s$ is of order $\mathcal{O}(|\mu|^{-2})$ in the original spatial variable. We neglect this boundary-layer correction. The constant $a_1$ is then determined from Eq.~\eqref{eq: condition 1d ss}, yielding
\[
a_1=\frac{2}{3}+\mathcal{O}(|\mu|^{-2})\,.
\]
We therefore arrive at the approximation
\begin{equation*}
\rho_s(x)=\frac{2}{3}\left[\cos\left(\sqrt{|\mu|}x\right)+1\right]
+\mathcal{O}(|\mu|^{-2}),
\qquad |x|\le \ell\,.
\end{equation*}

\underline{\emph{Steady states in 2D}. }  
We seek radially symmetric steady states \(\rho_s(\mathbf{x})=\rho_s(r)\), \(r=|\mathbf{x}|\), supported on \([0,\ell]\) with \(\rho_s(\ell)=\rho_s'(\ell)=0\) and regularity at \(r=0\). In radial coordinates the steady state equation is
\[
\frac{1}{r}\left[r\,\rho_s
\left(\rho_s'' + \frac{1}{r}\rho_s' + |\mu|\rho_s\right)'\right]'
= \rho_s(1-\rho_s).
\]

For strong adhesion (\(|\mu|\gg1\)) the aggregate is localised on the length scale \(r\sim|\mu|^{-1/2}\). Introducing $r={\hat r}/{\sqrt{|\mu|}}$,
the leading-order equation (dropping the \( \mathcal{O}(|\mu|^{-2})\) on the right-hand side) together with the boundary condition gives
\[
\rho_{\mathrm{out}}'' + \frac{1}{\hat r}\rho_{\mathrm{out}}' + \rho_{\mathrm{out}} = C_1,
\]
whose radially symmetric solution is
\[
\rho_{\mathrm{out}}(\hat r) = a_1 J_0(\hat r) + a_2,
\]
with \(J_0\) the Bessel function of the first kind. Requiring compact support, $\rho_{\mathrm{out}}(\hat{r}) = 0$ at $\hat{r} = j_{1,1}$, fixes $a_2 = -a_1\,J_0(j_{1,1})$ and the support radius is determined by the first positive zero of \(J_1\),
\[
\ell=\frac{j_{1,1}}{\sqrt{|\mu|}},
\]
where \(j_{1,1}\) is the first positive root of \(J_1\). Hence, in the original radial variable the leading-order approximation is
\begin{equation}
\label{eq:rho2d_leading}
\rho_s(r)=a_1\!\left[1 + \frac{J_0\left(\sqrt{|\mu|}\,r\right)}{|J_0(j_{1,1})|}\right]
+\mathcal{O}(|\mu|^{-2}),\qquad r\in[0, \ell].
\end{equation}

Proliferation produces a narrow boundary-layer near \(r=\ell\); its contribution to \(\rho_s\) in the original spatial variable is of order \(\mathcal{O}(|\mu|^{-2})\) and is therefore neglected given the accuracy of Eq.~\eqref{eq:rho2d_leading}. The amplitude \(a_1\) is fixed by Eq.~\eqref{eq: condition 2d ss}, which yields $a_1 = 1/{2} +\mathcal{O}(|\mu|^{-2})$. Thus the steady-state profile up to \(\mathcal{O}(|\mu|^{-2})\) is given by Eq.~\eqref{eq:rho2d_leading} with \(\ell=j_{1,1}/\sqrt{|\mu|}\). The maximum density corresponds to $r = 0$, which gives $\rho_\mathrm{max}\sim 1.74$.

\underline{\emph{Characteristic wavelength of the pattern}. } Following the linear stability analysis in Section~\ref{sec:linear-stability}, perturbations to the homogeneous steady state $\rho = 1$, with wavelength $\lambda$ will grow at rate
\begin{equation*}
    \lambda(k; 1) = -1 - \mu k^2 - k^4.
\end{equation*}

\noindent This expression depends only on the magnitude of the wavevector $k$, indicating that the instability is isotropic. 
Maximising $\lambda(k)$ with respect to $k$ gives
\begin{equation}
    k^{*} = \sqrt{\frac{|\mu|}{2}}, 
\end{equation}
\noindent corresponding to the most rapidly growing mode, with associated linear growth rate
\begin{equation*}
    \lambda(k^{*}) = -1 + \frac{\mu^{2}}{4}.
\end{equation*}
\noindent Therefore, the band of modes with $|\mathbf{k}| = k^{*}$ is the first to become unstable when $\mu < -2$. 
The corresponding characteristic pattern wavelength is
\begin{equation*}
    \ell_a = \frac{2\pi}{k^{*}}
    = 2\pi \sqrt{\frac{2}{|\mu|}}.
\end{equation*}
This wavelength defines the typical distance between cell aggregates.

\section{Density-dependent adhesion model}

\subsection{Travelling fronts}
\label{sec: travelling fronts dd}
We look for travelling fronts 
with speed $c = c(\mu) >0$ in the horizontal direction in the saturated adhesion model given by Eq.~\eqref{eq: local model density-dep}. We thus set $\rho(\mathbf{x},t) = \mathcal{P}(\xi)$ for $\xi = x-ct$, which yields
\begin{equation}
    c\mathcal{P}' - (\mathcal{P}(1-\mathcal{P})\mathcal{P}''')' +\mu (\mathcal{P}(1-\mathcal{P})\mathcal{P}')' + \alpha(\mathcal{P}^2\mathcal{P}')' + \mathcal{P}(1-\mathcal{P}) = 0\,\label{eq: tw density dep}\end{equation}
satisfying $\mathcal{P}(-\infty) = 1, \,\mathcal{P}(0) = 0$, assuming fronts are compactly supported. Recall that for this model, $\mathcal{P} = 1$ is always stable and $\mathcal{P} = 0$ is unstable, which are necessary conditions for the existence of a travelling wave solution.

In contrast with the unsaturated adhesion model, now the model also degenerates at $\mathcal{P}=1$, and hence we avoid studying the linearisation  near this state, as we expect nonlinear contributions that will not be captured by this method. We limit ourselves to the study of the limits: (i) $\alpha\rightarrow + \infty$, and $\omega=\mathcal{O}(1)$ (weak adhesion); and (ii) $\mu\rightarrow -\infty$ and $\alpha = \mathcal{O}(1)$ (strong adhesion). We show that we recover monotonic invasion fronts in both cases, as suggested by numerical solutions of the time-dependent model.

\subsection{Weak adhesion: fast monotonic invasion fronts}
\label{sec: weak adhesion saturated}
Here we write $\mu = \alpha-\omega$, with $\alpha\rightarrow+\infty$, and $\omega = \mathcal{O}(1)$. We rewrite the model in terms of $\omega$ which is fixed, so that
\begin{equation}
     c\mathcal{P}' - (\mathcal{P}(1-\mathcal{P})\mathcal{P}''')' +\mu (\mathcal{P}\mathcal{P}')' + \omega(\mathcal{P}^2\mathcal{P}')' + \mathcal{P}(1-\mathcal{P}) = 0\,.
\end{equation}
In particular, the outer solution will be, at leading order, identical to that of the unsaturated adhesion model. This is, when $\xi = \mathcal{O}(\sqrt{\mu})$, we expect the well-known Porous-Fisher solution, with a leading order wave speed $c\sim\sqrt{\mu/2}$. Higher-order contributions to the speed can be, in principle, determined analogously to Section~\ref{sec: weak adhesion constant}, and these will show differences with respect the unsaturated adhesion model. Similarly, the inner solution is also unaffected with respect to the unsaturated adhesion model. This is because additional terms in this model have a higher degeneracy at $\mathcal{P} = 0$, and thus only enter at higher orders in the asymptotic expansion.

\subsection{Strong adhesion: slow monotonic invasion fronts}\label{sec: strong adhesion saturated} Now we have $\mu = \alpha-\omega$, with $\mu\rightarrow-\infty$, and $\alpha = \mathcal{O}(1)$. In this limit we also expect $c$ to be decreasing with $|\mu|$. For now we write $c \leq \mathcal{O}(1)$, as the correct order for $c$ can be determined by using the leading order solution of the wave and the expression
\begin{equation}
    c = \int_{-\infty}^{+\infty}\mathcal{P}(\xi)(1-\mathcal{P}(\xi))\,\mathrm{d}\xi\,,\label{eq: speed frm tw}
\end{equation}
which is obtained by integrating Eq.~\eqref{eq: tw density dep}.

Eq.~\eqref{eq: tw density dep} degenerates at the back and at the front of the wave. Hence in this regime we could expect an invasion front which transitions from $\mathcal{P} = 1$ to $\mathcal{P} = 0$ in a finite length of width $\delta$. This is confirmed by numerical simulations where we also see that this corresponds to a thin region. Hence we restrict the analysis to an \emph{inner} layer centered at $\xi = -\delta/2$. We define $\xi = -\delta/2 + \zeta/\sqrt{|\mu|}$, so that with respect to $\zeta$ we have $\mathcal{P}(\zeta = 0) = 1/2$, and Eq.~\eqref{eq: tw density dep} becomes
\begin{equation*}
    c\sqrt{|\mu|}\mathcal{P}' - |\mu|^2(\mathcal{P}(1-\mathcal{P})\mathcal{P}''')' -|\mu|^2 (\mathcal{P}(1-\mathcal{P})\mathcal{P}')' + \alpha|\mu|(\mathcal{P}^2\mathcal{P}')' + \mathcal{P}(1-\mathcal{P}) = 0\,,
\end{equation*}
so the leading order solution satisfies
\begin{equation*}
    (\mathcal{P}_0(1-\mathcal{P}_0)\mathcal{P}_0''')' +(\mathcal{P}_0(1-\mathcal{P}_0)\mathcal{P}_0')' = 0\,.
\end{equation*}
This equation can be integrated once, and using boundary conditions we obtain
\begin{equation*}
    \mathcal{P}_0''' + \mathcal{P}_0' = 0\,,
\end{equation*}
which can be solved explicitly to give
\begin{equation*}
    \mathcal{P}_0(\zeta) = \frac{1}{2}\left(1 - \sin(\zeta)\right)\,,
\end{equation*}
where we used $\mathcal{P}_0 = 1$ when $\zeta = -\delta\sqrt{|\mu|}/2$, and $\mathcal{P}_0 = 0$, when $\zeta = \delta\sqrt{|\mu|}/2$.
With respect to the original variable, $\xi$, we have that $\delta = \pi/\sqrt{|\mu|}$, and 
\begin{equation*}
    \mathcal{P}(\xi) \sim \frac{1}{2}\left[1-\sin\left(\sqrt{|\mu|}\left(\xi + \frac{\delta}{2}\right)\right)\right] = \sin^2\left(\frac{\sqrt{|\mu|}\xi}{2}\right)\,,
\end{equation*}
for $\xi\in[-\pi/\sqrt{|\mu|},0]$. Using Eq.~\eqref{eq: speed frm tw} we obtain the asymptotic behaviour for the wave speed
\begin{equation*}
    c\sim\int_{-\pi/\sqrt{|\mu|}}^0 \,\sin^2\left(\frac{\sqrt{|\mu|}\xi}{2}\right) \cos^2\left(\frac{\sqrt{|\mu|}\xi}{2}\right)\,\mathrm{d}\xi = \frac{\pi}{8\sqrt{|\mu|}}.
\end{equation*}
\section{Numerical scheme}
\label{sec: finite volume}
We take advantage of the energy formulation of the model without proliferation and of the explicit solution for the logistic growth and use a splitting approach to solve the model numerically. For any time interval $[t, t+\delta t]$, we set $\tau = \delta t/2$ and split the equation into a conservative and a non-conservative part. These are then solved separately on successive time intervals. In Appendix~\ref{sec:appendix-energy} we wrote the model as 
\begin{equation*}
    \partial_t \rho = \nabla\cdot\left[m(\rho)\,\nabla\left(h(\rho)-\Delta\rho\right)\right] + \rho(1-\rho)\, ,
\end{equation*}
with 
\begin{equation*}
    h'(\rho) = \frac{d(\rho)}{m(\rho)} -\omega\,. 
\end{equation*}
For the two models used here, we can choose
\begin{equation*}
    h_{(\mathrm{I})}(\rho) =  \mu\rho,\quad h_{(\mathrm{II})}(\rho) = -\alpha\log(1-\rho) - \omega\rho\,.
\end{equation*}
Now we consider the conservative part
\begin{equation}
\partial_t\rho = \nabla\cdot\left[m(\rho)\,\nabla\left(h(\rho)-\Delta\rho\right)\right],\quad \rho(\mathbf{x},0) = \rho_{\text{in}}(\mathbf{x})\label{eq: conservative app}\tag{$\mathrm{C}$},
\end{equation}
and the non-conservative problem
\begin{equation*}
\partial_t\rho = \rho(1-\rho),\quad \rho(\mathbf{x},0) = \rho_{\text{in}}(\mathbf{x})\,.
\end{equation*}
Given the solution at time $t$, we write $\rho(\mathbf{x},t) :=\rho_t$ and approximate the solution at time $t+\delta t$ as:
\begin{equation*}
\rho_{t+\delta t}\approx (\mathcal{L}_{\tau/2}\circ \Phi_\tau\circ\mathcal{L}_{\tau/2})(\rho_t)\, ,
\end{equation*}
where $\mathcal{L}_t(\rho)$ is the solution of the logistic growth equation at time $t$ with initial data $\rho$; and $\Phi_t(\rho)$ is a numerical approximation at time $t$ and with initial data $\rho$ from the problem given by Eq.~\eqref{eq: conservative app}. This will be a good approximation if $\delta t$ is small enough and if $\Phi_t$ is a good approximation of the solution to the conservative problem. Moreover, we know that the map $\mathcal{L}_t$ is explicit and is given by
\begin{equation}
\mathcal{L}_t(\rho) = \frac{\rho}{ (1 - \rho) e^{-t} + \rho}\,.
\end{equation}
Below we detail an approach to approximating the conservative problem.

\subsection{Numerical solution of the conservative problem}
We use a finite-volume scheme for Eq.~\eqref{eq: conservative app}, based on the approach developed by Bailo et al.~\cite{bailo2021unconditional}. We present it in one spatial dimension for simplicity. The spatial domain $[0,L]$ is discretised into $N$ cells $C_i=[x_i,x_{i+1}]$, of length $\delta x$, and with centres at $x_{i+1/2} = (i+1/2) \delta x$, for $i = 0,\ldots,N-1$. We take $x_0 = 0$ and $x_N = L$. By integrating Eq.~\eqref{eq: conservative app} over $C_i$ we obtain
\begin{equation*}
\frac{\mathrm{d}}{\mathrm{d}t}\int_{C_i} \rho = \left[m(\rho) \partial_x\mathrm{w}\right]_{x_{i}}^{x_{i+1}}\,,\quad\mathrm{w} = h(\rho)-\partial_x^2\rho\,.
\end{equation*}
At cell centres, we approximate $\rho_i\approx  (\delta x)^{-1}\int_{C_i}\rho$. For $\mathrm{w}$ we use a centred difference for the Laplacian
\begin{equation*}
\mathrm{w}_{i+1/2}\approx h(\rho_{i+1/2}) - \frac{\rho_{i+3/2} + \rho_{i-1/2} - 2\rho_{i+1/2}}{\delta x^2}\,,\quad i = 0,\ldots,N-1\,,
\end{equation*}
where periodic boundary conditions are implemented by identifying $x_{-1/2} = x_{N-1/2}$ and $x_{N+1/2} = x_{1/2}$. Next we need to approximate the values of the flux $\mathrm{j} = m(\rho)\partial_x \mathrm{w}$ at the cell edges. We first approximate the velocity $\mathrm{v}=\partial_x \mathrm{w}$ using the finite difference:
\begin{equation*}
   %\mathrm{v}_i =  (\partial_x \mathrm{w})_{i}\approx \frac{1}{2}\left[\frac{\mathrm{w}_{i+1/2}-\mathrm{w}_{i-3/2}}{2\delta x}+\frac{\mathrm{w}_{i+3/2} - \mathrm{w}_{i-1/2}}{2\delta x}\right]\,,\quad i = 0,\ldots, N\,,
   \mathrm{v}_i =  (\partial_x \mathrm{w})_{i}\approx\frac{\mathrm{w}_{i+1/2}-\mathrm{w}_{i-1/2}}{\delta x}\,,
\end{equation*}
where, again, we use periodic boundary conditions. %and identify $x_{N+3/2} = x_{3/2}$ and $x_{-3/2} = x_{N-3/2}$. 
The flux is constructed using an upwind approximation
\begin{equation}
\mathrm{j}_i\approx  m(\rho_{i-1/2})(\mathrm{v}_{i})^+ + m(\rho_{i+1/2})(\mathrm{v}_i)^-\,,
\end{equation}
where $(x)^+ = \max(x,0)$ and $(x)^- = \min(x,0)$.
Finally, we solve the system of ordinary differential equations
\begin{equation*}
    \frac{\mathrm{d}\rho_i}{\mathrm{d}t} = \frac{\mathrm{j}_{i+1}-\mathrm{j}_i}{\delta x}\,,\quad i = 0,\ldots,N-1\,,
\end{equation*}
using the LSODA algorithm---implemented in \texttt{scipy.integrate}---which automatically switches between a nonstiff Adams method and an implicit BDF method.

\section{Parameters for main text figures}\label{sec: parameters}
\begin{itemize}
    \item Fig.~\ref{fig: fig 2}A shows a two-dimensional simulation on $[-L/2,L/2]^2$, with $L = 20$, $\delta x = 0.2$, $\delta t = 10^{-4}$, $\mu = -2$ with periodic boundary conditions. The initial condition is $\rho(\mathbf{x},0) = 1 + \mathrm{U}_{\mathbf{x}}$, where $U_\mathbf{x}$ is Uniform$(-0.1,0.1)$. Numerical solution is shown at times: $t = 3.25,\,5.75,\,50$.
    \item Fig.~\ref{fig: fig 3}A shows a numerical solution on a two-dimensional box $[-L_x/2,L_x/2]\times[-L_y/2,L_y/2]$, with $L_x =40$, $L_y = 20$, and $\delta x = 0.2$, $\delta t = 10^{-4}$, $\mu = 2$. The initial condition is $\rho(\mathbf{x},0) = \mathbb{1}_{\{|x|< 3\}}\left[1 + 0.1\cos(\pi y/10)\right]$. Numerical solution is shown at times: $t = 0,\,10,\,20$.
   \item For Fig.~\ref{fig: fig 3}B we simulated the unsaturated adhesion model in one spatial dimension, starting from the initial condition $\rho(x,0) = \mathbb{1}_{|x|<...}$ and letting the travelling wave evolve in the positive and negative axis. We used a domain $[-L/2,L/2]$ with $L$ large enough so that boundary conditions have no influence, and with $\delta x = 0.1$ and $\delta t = 10^{-4}$. We then estimated the wave speed using the trapezoidal rule applied to
   \begin{equation}
       c = \frac{1}{2}\int_{-L/2}^{L/2}\rho(\xi)(1-\rho(\xi))\,\mathrm{d}\xi\,,\label{eq: wave speed numerical}
   \end{equation}
   where the factor $1/2$ accounts for the bidirectional propagating front.
   \item Fig.~\ref{fig: fig 3}C-D show snapshots of the one-dimensional travelling waves but restricted to $x > 0$. We set the origin $\xi = 0$
    to match the location of the front $\rho(0) \sim 10^{-6}$.
    \item Fig.~\ref{fig: fig 3}E shows a numerical solution on a two-dimensional box $[-L/2,L/2]^2$, with $L =40$, and $\delta x = 0.2$, $\delta t = 10^{-4}$, $\mu = -1.5$. The initial condition is $\rho(\mathbf{x},0) = \mathbb{1}_{\{|x|< 4 + 0.2\cos(\pi y/5)\}}$. Numerical solution shown at times: $t = 0,\,100,\,200$.
     \item Fig.~\ref{fig: fig 4}A shows a numerical solution on a two-dimensional box $[-L/2,L/2]^2$, with $L =40$, and $\delta x = 0.2$, $\delta t = 10^{-4}$, $\mu = -4$. The initial condition is $\rho(\mathbf{x},0) = \mathbb{1}_{\{|x|< 4 + 0.2\cos(\pi y/5)\}}$. Numerical solution is shown at times: $t = 0,\,75,\,250$.
       \item Fig.~\ref{fig: fig 4}C shows results from numerical simulations on a two-dimensional box $[-L/2,L/2]^2$, with $L =8$, and $\delta x = 0.2$, $\delta t = 10^{-4}$, $\mu = -4$. The initial condition is $\rho(\mathbf{x},0) = \mathbb{1}_{\{|\mathbf{x}|< 2\}}$ and the equation is solved until convergence to a steady state.
       \item Fig.~\ref{fig: fig 5} shows numerical simulations in a one-dimensional box with periodic boundary conditions with $L = 60$, and $\delta x = 0.05$, $\delta t = 10^{-5}$. Solutions are shown at $t = 60$ and wave speeds are estimated using Eq.~\eqref{eq: wave speed numerical}. 
    \end{itemize}

\newpage

\end{document}